\documentclass[conference,compsoc]{IEEEtran}
\usepackage{listings}
\usepackage[table,xcdraw]{xcolor}
\usepackage{multirow}
\usepackage{diagbox}
\usepackage{enumitem}
\usepackage{comment}
\usepackage{soul}
\usepackage{url}
\usepackage{xcolor}
\usepackage{multirow}
\usepackage{hyperref}
\usepackage{microtype}
\usepackage{listings}
\usepackage{xcolor}
\usepackage{array}
\usepackage{graphicx}
\ifCLASSOPTIONcompsoc
  \usepackage[nocompress]{cite}
\else
  \usepackage{cite}
\fi

\usepackage{amsmath}
\usepackage{algorithmic}
\usepackage{array}
\newif\ifrevision
\newcommand{\textrevision}[1]{%
  \ifrevision
    \textcolor{blue}{#1}%
  \else
    #1%
  \fi
}

\begin{document}

\title{Vulnerable Models in the AI Supply Chain: An Empirical Study of Adoption, Propagation, and Usage Patterns}
\author{Anonymous Authors}

\author{\IEEEauthorblockN{The Anh Nguyen\IEEEauthorrefmark{1},
Triet Huynh Minh Le\IEEEauthorrefmark{1},
M. Ali Babar\IEEEauthorrefmark{1}\IEEEauthorrefmark{2}}
\IEEEauthorblockA{\IEEEauthorrefmark{1}School of Computer Science and Information Technology, Adelaide University, Adelaide, Australia\\\{theanh.nguyen, triet.h.le, ali.babar\}@adelaide.edu.au}
\IEEEauthorblockA{\IEEEauthorrefmark{2}ElevexAI Systems}
}

\maketitle
\begin{abstract}
The rapid integration of Artificial Intelligence (AI) models has established a complex supply chain, but many of these shared artifacts harbor severe vulnerabilities such as insecure serialization and architectural backdoors. While prior efforts have extensively scanned model registries to flag malicious artifacts, a critical gap remains in understanding how developers actually adopt and integrate these unsafe models into downstream software systems. We bridge this gap by conducting a large-scale empirical study on the usage and propagation of 16,977 known vulnerable models on HuggingFace flagged by InsightDB. We track their prevalence across GitHub repositories, software packages, fine-tuning chains, and deployed applications. Our analysis reveals that, despite being publicly flagged, these vulnerable models appear in over 21,000 code locations across more than 5,000 open-source repositories, with 3,757 distinct flagged models (22.1\%) actively referenced in at least one project. To understand how these models are integrated in code and identify recurring usage patterns, we categorize and analyze the recurring patterns surrounding them via a customized pipeline that combines Code Property Graphs extraction with a tailored Frequent Subgraph Mining approach. Across all patterns, we find no evidence of provenance verification or model sanitization logic, with developers frequently bypassing security parameters or delegating trust to permissive framework defaults. These findings underscore that scanning model files alone is insufficient, highlighting the urgent need for context-aware security analysis and stricter integration guardrails at the application layer of the AI supply chain.

\end{abstract}

\section{Introduction}
The proliferation of Artificial Intelligence (AI) models has given rise to a complex, multi-layered AI supply chain.
Central to this ecosystem are model registries, e.g., Hugging Face, which host millions of pre-trained models that developers can easily download and integrate into open-source frameworks, enterprise tools, and user-facing applications.
This widespread reuse accelerates innovation but introduces broad exposure surfaces where the security of a downstream application is highly dependent on the integrity of its upstream AI components~\cite{jiang2022empirical, meiklejohn2025machine, hu2025large}.
This reliance on third-party artifacts introduces significant security risks.
Many pre-trained models are vulnerable to threats such as arbitrary code execution via insecure serialization formats (e.g., Python's pickle) or embedded architectural backdoors.
While hosting platforms and security advisories have begun employing scanners to flag these vulnerable artifacts~\cite{huggingface_protectai_security, protectai_insights, zhao2024models}, the actual downstream impact of these vulnerabilities depends on whether and how developers consume the vulnerable models in their own codebases~\cite{digregorio2026security}.
For instance, a model flagged for insecure deserialization poses little practical risk if never loaded, but becomes a concrete threat when a developer passes it directly to a loading API with permissive security settings, a scenario we find to be common in practice (Section~\ref{subsec.res.rq2}).

Current research on AI supply chain security has primarily focused on identifying vulnerabilities within model files themselves~\cite{kellas2025pickleball, liu2026art, casey2024large} or quantifying their presence on model hosting platforms~\cite{zhao2024models, siddiq2026empirical}.
While these analyses are essential for threat detection, they offer limited insights into the practical realities of downstream integration~\cite{digregorio2026security}.
Specifically, it remains largely unknown to what extent developers adopt vulnerable models, how these models propagate through the broader software ecosystem, and what recurring coding patterns are used to integrate them.
This gap hinders the development of effective, context-aware mitigation for the application layer of the AI supply chain.

We bridge this gap by extending the analytical focus beyond scanning only model artifacts to include developer usage.
We conduct a comprehensive empirical investigation of the adoption, propagation, and integration patterns of 16,977 known vulnerable models on Hugging Face (HF).
First, we measure the exposure of these vulnerable models across four dimensions of the AI supply chain: direct source code references on GitHub (GH), software package distributions, upstream and downstream fine-tuning chains, and deployed HF Spaces.
Second, to understand how these models are integrated in developers' source code, we design a pipeline that combines Code Property Graph (CPG) extraction with custom normalization and constrained Frequent Subgraph Mining (FSM) using the Top-K Graph (TKG) algorithm.
This approach enables us to systematically discover and categorize the recurring structural motifs of vulnerable model usage. 
Our key \textbf{contributions} are as follows:
\begin{itemize}[]
\item We reveal the widespread adoption of vulnerable models, despite public security advisories, with over 21,000 code occurrences in more than 5,000 repositories, and further propagation through 419 fine-tuned derivatives, 185 software packages, and over 1,200 deployed applications.
\item We introduce a data-driven approach that leverages FSM via the TKG algorithm to automatically surface recurring propagation structures and candidate sinks directly from code. This approach complements traditional manual CPG workflows by reducing the manual effort required for source-sink specification, providing an empirically grounded foundation for future vulnerability query design.
\item We show that vulnerable model usages recur in a small number of patterns and organize them into four motifs: Direct Usage, Local Propagation, Method Propagation, and Condition-Guarded Propagation, deriving a compact taxonomy of vulnerable model integration structures.
\item Crucially, we observe a systemic lack of model sanitization or cryptographic provenance verification across all extracted patterns, demonstrating that developers consistently delegate security to permissive framework defaults or explicitly bypass available safeguards during model integration.
Our findings show that artifact-level model scanning, while useful, is insufficient on its own to secure the AI supply chain, as security risks persist and amplify through downstream code integrations. 
We provide empirical evidence and concrete structural insights that can inform the design of context-aware static analysis tools, stronger secure-by-default framework behaviors, and integration-layer guardrails to reduce the gap between artifact-level detection and application-level protection.
\end{itemize}

\section{Related Work and Motivation}
\label{sec.related}

\subsection{Security Threats in AI models}
The vulnerability of machine learning models has been primarily associated with model artifacts, e.g., insecure serialization formats and model backdoors.
Python's \texttt{pickle} module, the underlying loading method of PyTorch, allows arbitrary code execution during deserialization~\cite{python_pickle_module}, making it a recognized supply chain risk.
Kellas et al.~\cite{kellas2025pickleball} characterize this attack and propose a framework that analyzes the deserialization process to restrict potentially harmful executables.
Liu et al.~\cite{liu2026art} show that stealthy pickle payloads can evade current detection tools, e.g., PickleScan~\cite{drewt2025pytorch, newspicklescan}.
Zhao et al.~\cite{zhao2024models} leverage control and data flow queries on source-sink relationships to detect malicious models.
Their findings demonstrate that scanning model files alone is insufficient to capture the full risk of unsafe model loading.
Zhu et al.~\cite{zhu2025my} demonstrate architectural backdoors via TensorFlow APIs, where malicious behavior is embedded directly in the computation rather than in serialized artifacts.
Li et al.~\cite{li2021deeppayload} propose \textsc{DeepPayload}, injecting backdoors into deployed deep learning models on mobile devices and showing that the attack surface extends beyond model files to the model architectures.
Xiao et al.~\cite{xiao2018security} provide a characterization of security risks in deep learning implementations such as operator-level vulnerabilities.
More recently, Ding et al.~\cite{ding2025rusty} detect malicious configurations that trigger code execution during model loading without modifying the model artifact itself, while Li et al.~\cite{li2025exploring} explore typosquatting threats where malicious repositories masquerade as popular models.

\begin{figure}
\centering
\includegraphics[width=\linewidth]{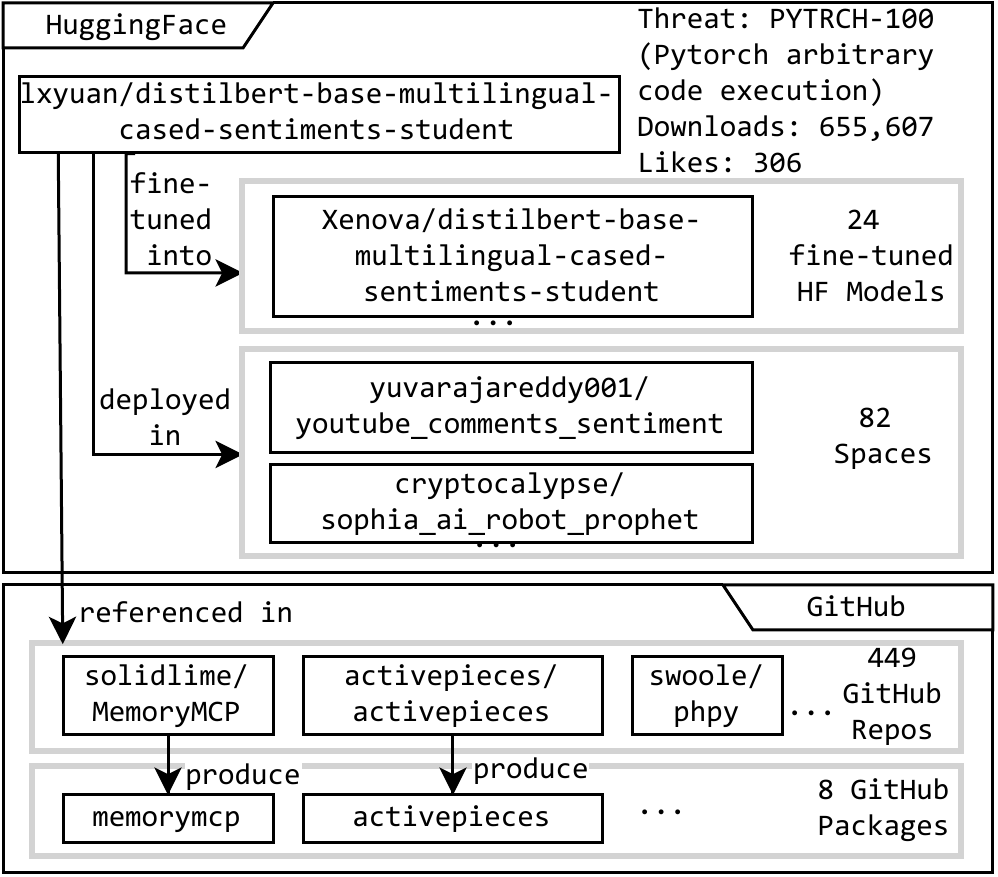}
\caption{Example of propagation of a vulnerable model.}
\label{fig:propagation_example}
\end{figure}

Several studies also measure the prevalence of these threats on model hosting platforms.
Casey et al.~\cite{casey2024large} analyze model files to determine whether malicious payloads are triggered during loading workflows.
Siddiq et al.~\cite{siddiq2026empirical} study remote code execution risks across hosting ecosystems, finding that unsafe serialization formats remain prevalent despite the availability of safer alternatives~\cite{trailofbitspublications}.
Di Gregorio et al.~\cite{digregorio2026security} study security posture more broadly, showing that existing mitigation mechanisms often provide only partial protection and frequently shift responsibility to users.
While these works characterize the threats present in model artifacts and hosting platforms, they offer limited insights into the security of the AI supply chain where these distinct threats actively coexist, and how downstream developers actually integrate these vulnerable models.
For example, the sentiment classifier model \texttt{lxyuan/distilbert-student}~\cite{lxyuan2023distilbert-multilingual-sentiment} has accumulated over 655K downloads~(Figure~\ref{fig:propagation_example}).
Despite being flagged for arbitrary code execution via \texttt{pickle} deserialization~\cite{protectai_knowledge_base, protectai2024distilbert-sentiment-overview}, the model appears in 594~references across 449~repositories.
The risk amplifies through propagation with 24 models fine-tuned from this base, some inheriting the vulnerable serialization format, while also being deployed in 82 HF Spaces that are exposed directly to end users.
Five repositories even ship production container images containing this model.
This example demonstrates that scanning model files alone cannot capture the practical severity of a vulnerable model, and the risk extends to how developers integrate and propagate the model in their own codebases.
Our work complements the current literature by systematically investigating both the scale of vulnerable model adoption, propagation, and the recurring coding patterns through which these models are consumed. 
We provides the empirical foundation needed to design more effective detection and context-aware mitigation strategies to further secure the application layers of the AI supply chain.

\subsection{The AI Model Supply Chain Ecosystem}
The rapid growth of model registries, particularly HF with more than two million models \cite{huggingface_hub_docs}, has created a model supply chain that increasingly resembles the software package ecosystem.
Jones et al.~\cite{jones2024what} provide a systematic review of HF, validating the dominance of the platform for pre-trained model sharing.
Jiang et al.~\cite{jiang2022empirical} document trust assumptions developers make when consuming model artifacts and their security risks in the pre-trained model supply chain.
The PeaTMOSS dataset~\cite{jiang2024peatmoss} further enables large-scale mining of pre-trained model reuse in open-source software.
Recent work has also begun to frame these ecosystems more explicitly as supply chains.
Hu et al.~\cite{hu2025large} identify provenance, integrity verification, and dependency management as key unresolved issues in the LLM supply chain.
Huang et al.~\cite{huang2025lifting} examine the composition and risks of the LLM supply chain, finding that model dependencies are often implicit and poorly tracked.
Meiklejohn et al.~\cite{meiklejohn2025machine} argue that ML models face a supply chain problem analogous to software packages and propose signing and provenance attestation as countermeasures.
Yasmin et al.~\cite{yasmin2026software} study the reuse and integration of pre-trained models in open-source projects, treating models as a new form of software dependency.
Liu et al.~\cite{liu2025empirical} show that LLM repositories frequently depend on packages with known vulnerabilities, indicating that software and model supply chain risks can compound.
Patel et al.~\cite{patel2026towards} survey attacks and mitigations across the MLOps lifecycle, spanning data, model, and deployment stages.
Prior studies have analyzed security issues discussed by developers in broader software-development communities~\cite{le2020puminer,le2021large}, while Nguyen et al.~\cite{nguyen2026securing} analyze developer-reported security issues in AI projects and find that supply chain concerns are a recurring theme in developer discussions.
As the central distributor in this ecosystem, HF began to incorporate security-oriented governance mechanisms.
The platform integrates third-party advisory signals via ProtectAI and JFrog~\cite{huggingface_pai_2025, huggingface_protectai_security}.
Protect AI's Insights DB aggregates model security issues across deserialization, architectural, and runtime threats~\cite{protectai_insights, protectai_knowledge_base}, derived from actively scanning the model hub.
These scanning and advisory mechanisms provide valuable sources of intelligence that can serve as a foundation for studying how vulnerable models propagate through and are integrated into downstream software.
However, to the best of our knowledge, no existing study has leveraged these sources to systematically analyze the adoption, propagation, and model usage patterns in real developer codebases.

\subsection{AI Vulnerabilities Analysis Techniques}
The modern AI development lifecycle forms a multi-layered supply chain in which foundation models are integrated through open-source frameworks, web applications, and APIs.
Code-centric security analyses operate directly on source-code representations~\cite{le2022towards,le2022survey,nguyen2024automated}; for downstream projects, a common starting point is the Abstract Syntax Tree (AST)~\cite{le2020deep}.
ASTs convert source code into an organized, hierarchical tree structure, allowing for the analysis of source code syntax.
However, while ASTs excel at representing the strict syntactic structure of a program, they lack semantic depth.
Specifically, AST does not fully capture the data-flow and control-flow dependencies required to trace how a vulnerable model identifier propagates after it is loaded.
To overcome these limitations, a Code Property Graph (CPG) \cite{yamaguchi2014modeling, joern_cpg_spec} can be utilized. 
CPG provides a comprehensive, joint representation that combines AST with a Control Flow Graph (CFG) and a Program Dependence Graph (PDG) \cite{joern_cpg_spec}, in which Control Dependence Graph (CDG) and reaching definition (REACHING\_DEF) edges encode conditional and data flows.
This unified structure effectively models both the syntax and the semantic data flows and control structures of the target application~\cite{le2022use}.
By constructing CPGs of downstream projects, we can accurately track the lineage, usage context, and sanitization pathways of integrated AI models.

While graph queries and predefined rules have been used to search graph representations for security vulnerabilities, e.g., MalHug \cite{zhao2024models} leverages CFG and data-flow-based queries on source-sink relationships, this approach exhibits certain limitations.
In particular, predefined queries can only detect patterns that have already been explicitly specified \cite{yamaguchi2014modeling, wi2022hiddencpg,le2021deepcva, khodayari2021jaw,li2022mining}.
To uncover how vulnerable models are used by developers in real-world settings, such patterns may not be known beforehand~\cite{le2019automated}.
This gap motivates a data-driven discovery approach that can surface recurring structures directly without requiring target sinks to be specified in advance.
Frequent Subgraph Mining (FSM)~\cite{yan2002gspan} is well-suited for this task, as it can automatically extract recurring structural patterns from large graph databases.
Related software security research has used this method to discover discriminative bug signatures \cite{cheng2009identifying},  security patch patterns \cite{song2019inferring}, and recent work has shown the value of automatic pattern mining for API misuses \cite{jiang2024app}.
However, to the best of our knowledge, FSM has not previously been applied to CPGs derived from AI model usage code or to discover how vulnerable model identifiers are integrated.
In this work, we design a tailored approach that adapts FSM through the Top-K Graph (TKG) algorithm~\cite{fournier-viger2019tkgd}, and combine it with custom CPG normalization and a root-node constraint mining strategy to ensure that all discovered patterns are anchored to vulnerable model identifiers (Section~\ref{sec.method}).
By applying TKG to CPGs, we can automatically identify, extract, and systematically discover recurring coding patterns during model integration.

\section{Research Questions}
\label{sec.rqs}
To understand vulnerable model adoption, propagation, and usage, we investigated two Research Questions (RQs):

\textbf{RQ1: How widespread are vulnerable model usages in the AI supply chain ecosystem?}
While HF security scanners can flag vulnerable model artifacts, the downstream impact of these models depends on whether they are actually adopted by developers.
This question measures the exposure surfaces of flagged models across four dimensions: direct code references, software packages, fine-tuning chains with upstream and downstream lineage, and deployed applications.

\textbf{RQ2: What recurring usage patterns are associated with vulnerable models, and what security risks these patterns imply?}
Knowing that vulnerable models appear in real projects is insufficient for mitigation without understanding their source code integrations. 
This question identifies recurring motifs through which vulnerable models are introduced, propagated, and consumed by sink APIs, and examines security-relevant risks these motifs may suggest.

\section{Methodology}

\label{sec.method}
\begin{figure*}
    \centering
    \includegraphics[width=\linewidth]{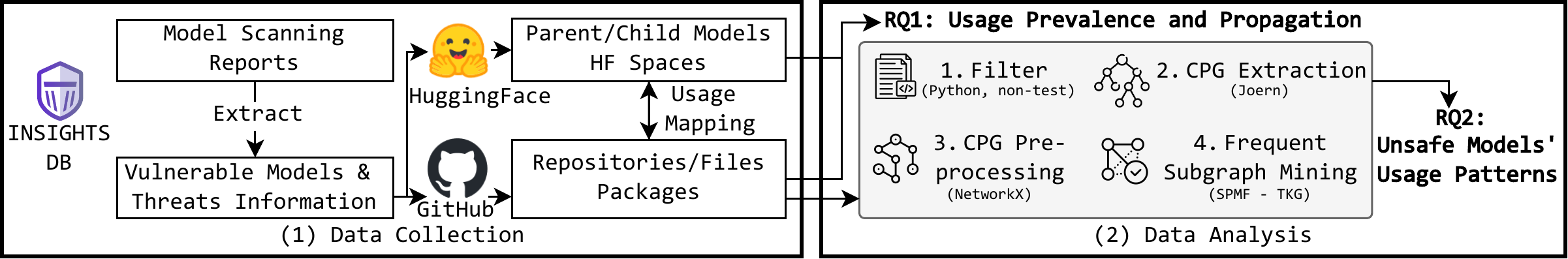}
    \caption{Overview of the research process.}
    \label{fig:method}
\end{figure*}

\begin{figure*}
    \centering
    \includegraphics[width=\linewidth]{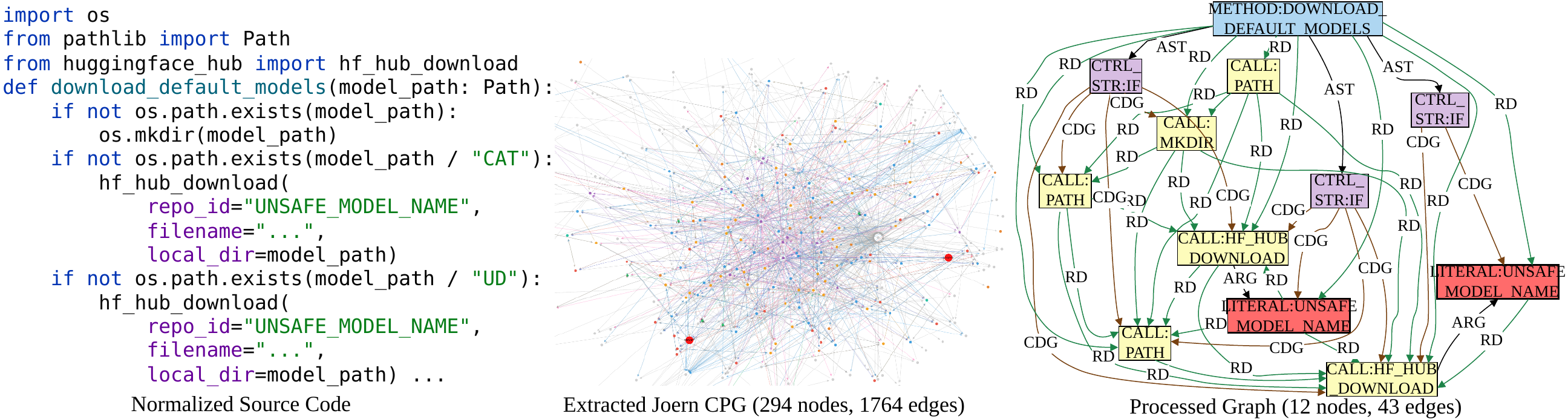}
    \caption{Example of the CPG Processing step for a sample script from the CAMeL-Lab/camel\_parser repository.}
    \label{fig:method_example}
\end{figure*}

\textbf{Overview.}
As depicted in Figure \ref{fig:method}, our methodology consists of two main steps: collection of vulnerable model data and data analysis.
First, we collected vulnerability reports from InsightDB and extracted the model and threat information, traced their respective parent/child models and GH repositories and packages.
For RQ1, we analyzed the extent to which vulnerable models are used and propagated in the ecosystem. 
For RQ2, we generated the CPG and employed FSM on the generated graphs to extract the common usage patterns of these vulnerable models.
All processing logic, dataset, and source code are available in our replication package~\cite{replication_package_vul_model_usage}.

\subsection{Data Collection}
\label{subsec.method.collect}
We started by curating a dataset of vulnerable AI models from the ProtectAI Insights DB \cite{protectai_insights}.
\textrevision{The database is officially integrated with HF~\cite{huggingface_protectai_security} and provides a curated source of model-centric security reports. 
By April 2025, it contained 4M scans across 1.4M repositories~\cite{huggingface_pai_2025}, indicating substantial ecosystem coverage.
Prior research also adopted ModelScan, InsightsDB's open-sourced scanner, to study model security due to its ability to detect security issues across various model formats~\cite{zhao2024models,liu2026art}.
As of January 2026, this dataset contains \textbf{16,977} unique model identifiers, with three main threats categories: deserialization, backdoor, and runtime threats.}
Deserialization threats occur when malicious code are embedded in serialized artifacts and executed during model loading.
Backdoor threats involves hidden architectural modifications that induce malicious or manipulated outputs under specific trigger inputs.
Runtime threats refer to unsafe operators or hidden execution logic that are triggered during model execution or inference.
The full threat descriptions are available in ProtectAI's knowledge base~\cite{protectai_knowledge_base}.

To obtain the usage of vulnerable models in the AI supply chain ecosystem, we utilized GH Code Search API to locate open-source repositories referencing these models.
Using the vulnerable model identifiers collected earlier as exact-match query strings, we queried for repositories and downloaded the complete raw source code for every file containing a match and their metadata.
Concurrently, to track the propagation of vulnerable models, we queried the HF Hub API for metadata tags to construct the fine-tuning chain.
We also queried the HF Spaces API to identify deployed applications that have source code references to these vulnerable models. 

\subsection{Data Analysis and Pattern Extraction}
\label{subsec.method.graph}
To answer the RQs, we performed different quantitative analysis on the collected dataset. 
For \textbf{RQ1}, to comprehensively quantify the risks posed by vulnerable models, we analyze the adoption and propagation of these models across four distinct dimensions of the AI supply chain: code references, package distribution, fine-tuning chains, and deployed applications. 
Each dimension captures a different way in which vulnerable models persist beyond their original repository page on the model hub.
First, we analyze \textit{code references} to measure direct integration of flagged models into downstream software projects.
A model that appears in source code has moved beyond passive availability on the hub and has become part of an application's implementation logic.
Second, we analyze \textit{package distribution} to capture cases where code referencing flagged models is redistributed as reusable software artifacts.
This surface is important because packages can propagate vulnerable models beyond the original repository and make unsafe integrations easier to inherit indirectly. 
We therefore identified package artifacts associated with matched repositories and recorded their occurrence as an additional distribution surface.
Third, we analyze \textit{fine-tuning chains} to measure how vulnerable models continue to propagate through the model lineage.
Vulnerable models may not only be referenced in their original forms, but may instead serve as the base for fine-tuned variants.
To capture this dimension, we extracted \texttt{base\_model} tags from the metadata~\cite{mitchell2019model, huggingface_model_cards}.
These tags link the models with their original versions, allowing for the reconstruction of relationships between vulnerable models and their upstream/downstream versions.
Lastly, we analyze \textit{deployed HF Spaces} to capture application-level exposure in publicly accessible model-powered applications.
While repository and package references show code integrations and software distributions, HF Spaces reveal where flagged models may already be embedded in user-facing applications.

To address \textbf{RQ2}, we investigated recurring code patterns associated with vulnerable models as follows:

\textbf{CPG Processing.}
To ensure that the analysis captured meaningful program semantics related to real-world AI model usage rather than testing or experimentation, we restricted our corpus to Python source files that were not test files.
Python was selected because it was not only the dominant language in our dataset but also the primary language used in most open-source AI and machine learning frameworks.
Test files were excluded to focus on production code paths where vulnerable models are actually integrated and executed.
This filtering resulted in a candidate set of \textbf{1,156} files across \textbf{817} repositories.
Then, to establish a precise and queryable anchor for our analysis, we replaced every occurrence of a vulnerable model identifier (Section~\ref{subsec.method.collect}) in each source file with the string \texttt{UNSAFE\_MODEL\_NAME} before graph generation.
The replacement targeted only string literals and did not alter program semantics, since model identifiers were used as literal values and not executed.
This allowed the unsafe source to appear as \verb|LITERAL:UNSAFE_MODEL_NAME| nodes, making it uniformly identifiable regardless of the original string.
CPGs were then generated using \textit{Joern}~\cite{joern_tool}, a static analysis framework that is widely used for security analysis \cite{wi2022hiddencpg, suneja2020learning, bowman2020vgraph}.
Since auto-generated CPGs were highly dense with hundreds of nodes for a simple function~\cite{yamaguchi2014modeling, joern_cpg_spec}, applying FSM directly to raw CPGs would be computationally expensive and semantically noisy.
To address this issue, we designed a normalization process, illustrated in Figure~\ref{fig:method_example}, specifically tailored to our analysis as follows:

\textbullet\ Nodes and Edges Filtering: To reduce graph density while retaining program semantics, we removed metadata and compiler nodes while retaining only semantically crucial nodes~\cite{joern_cpg_spec}: \texttt{IDENTIFIER} (\texttt{ID}), \texttt{CALL}, \texttt{LITERAL}, \texttt{METHOD}, \texttt{IMPORT}, \texttt{BLOCK}, \texttt{CONTROL\_STRUCTURE}, and edges: \texttt{AST}, \texttt{CFG}, \texttt{CDG}, \texttt{ARGUMENT} (\texttt{ARG}), \texttt{CALL}, \texttt{REACHING\_DEF} (\texttt{RD}).

\textbullet\ Propagation Subgraph Extraction: 
Joern provides a backward data-flow slicer (starting from sinks) and a usage slicer (describing intra-procedural variable interactions)~\cite{joern_cpg_slicing}, but neither supports forward slicing from a designated source node.
Since we needed to trace how the source node (\verb|LITERAL:UNSAFE_MODEL_NAME|) propagated both forward and backward in the source code, we designed a custom slicer over the exported CPG using NetworkX~\cite{networkx}.
Starting from unsafe sources, we (i)~traced forward along \texttt{RD} edges to collect all data-flow descendants;
(ii)~traced backward via \texttt{AST} edges to identify the enclosing methods and their AST trees; and
(iii)~applied one-hop expansion to extract the bodies of any invoked methods.
This yielded a semantically focused slice centered on the usage of the vulnerable model.

\textbullet\ Label Normalization: 
To ensure that structurally equivalent patterns across different codebases could be recognized as identical by FSM regardless of the naming conventions, we normalized node and edge labels.
Specifically, \texttt{CALL} and \texttt{METHOD} nodes were mapped to functional groups based on their \texttt{METHOD\_FULL\_NAME}:
Python operators, e.g., \texttt{COMPARE} or \texttt{ASSIGN}, to \texttt{CALL:OP\_*};
built-in functions, e.g., \texttt{ITER} or \texttt{NEXT}, to \texttt{CALL:BUILTIN\_*};
and framework and library calls, e.g., \texttt{torch} or \texttt{keras}, to \texttt{CALL:\-<\-MODULE>}.
\texttt{LITERAL} node labels were normalized to \texttt{TRUE} or \texttt{FALSE} if they were boolean values, as these reflected control flags in the source code; other literals were pruned as non-semantic (except for \texttt{LITERAL:UNSAFE\_MODEL\_NAME} nodes).
In the control flow, \texttt{CONTROL\_STRUCTURE} carried conditional information based on its \texttt{control\_structure\_type}, thus we normalized its labels using this attribute to yield labels such as \texttt{IF} or \texttt{TRY}.
Edge labels were retained as-is since edge types already carried precise semantic meanings defined by the CPG specification~\cite{joern_cpg_spec}.

\textbullet\ Graph Simplification: 
To further reduce noise without losing structural semantics, we removed intermediate nodes and duplicated edges.
While \texttt{BLOCK} and \texttt{ID} nodes are necessary during forward slicing to trace the propagation, once subgraphs are extracted these nodes appear so frequently across all slices that they acted as noise rather than discriminating structural features.
We therefore collapsed these nodes by rewiring edges directly between their neighbors and removed duplicated edges by retaining only the most semantically specific edge type between each pair of nodes, following the ordering $\texttt{ARG}>\texttt{CALL}>\texttt{RD}>\texttt{CDG}>\texttt{CFG}>\texttt{AST}$.
This order reflected decreasing semantic specificity, where direct data flow and method execution were prioritized.
\texttt{ARG} encodes direct parameter passing to a callee.
\texttt{CALL} represents method invocations.
\texttt{RD} captured data-flow dependency, where definition of nodes were reused at other nodes.
\texttt{CDG} encoded conditional executions, where execution of nodes was governed by control predicates.
In contrast, \texttt{CFG} captured execution orders, and \texttt{AST} carried pure syntactic relation.
Thus, \texttt{CFG}/\texttt{AST} edges are redundant if two nodes were already connected by more data-specific edges~\cite{joern_cpg_spec}.

Finally, as different ML frameworks exhibit distinct API usage conventions, aggregating scripts across frameworks risks obscuring framework-specific contexts in which vulnerable model names appear.
To preserve these distinctions and extract more fine-grained patterns, we split the normalized dataset by detected framework prior to mining.
Framework detection was performed by tracing the model name and the assigned variables through the call graph, identifying the invoked framework APIs, and matching the import strings of that call against a list of HF-compatible frameworks obtained from the HF documentation \cite{huggingface_docs}.
Repositories and files with no resolvable frameworks (e.g., using models  with no direct library calls) were grouped into the Other category (see Table~\ref{tab:framework_distribution}).
This splitting was further motivated by the uneven framework distribution in our final corpus, where patterns from the most-represented frameworks would otherwise dominate and suppress the recurring contexts unique to less-frequently used frameworks, consistent with prior findings that less-represented software settings can exhibit distinct characteristics that are obscured by high-resource settings~\cite{le2024software}.
The framework distribution of our normalized dataset is presented in Table~\ref{tab:framework_distribution}.

\begin{table}[t]
\centering
\caption{Distribution of Dataset by Framework}
\label{tab:framework_distribution}
\resizebox{\columnwidth}{!}{%
\begin{tabular}{l|r|r}
\textbf{Framework} & \textbf{\# Repositories} & \textbf{\# Files} \\ \hline
Transformers         & 332 & 470 \\
Hugging Face Hub (HF Hub)     & 168 & 178 \\
PyTorch              &  54 &  98 \\
Ultralytics          &  44 &  62 \\
TensorFlow           &  20 &  21 \\
ONNX                 &  11 &  15 \\
Sherpa               &  10 &  91 \\
Diffusers            &   9 &  19 \\
PEFT                 &   7 &  16 \\
Fastai               &   6 &   7 \\
Flair                &   6 &   6 \\
Joblib               &   5 &   7 \\
Keras                &   5 &   9 \\
Datasets             &   4 &   5 \\
Optimum              &   3 &   4 \\
SentenceTransformers (ST) &   3 &   6 \\
Other                & 127 & 142 \\ \hline
\textbf{Total}       & \textbf{817} & \textbf{1{,}156} \\
\end{tabular}
}
\end{table}

\textbf{Frequent Subgraph Mining.}  
To identify recurring patterns of vulnerable model usage, we customized the Top-K Frequent Subgraph (TKG) algorithm \cite{fournier2019tkg, fournier2016spmf}.
Foundational FSM algorithms, e.g., gSpan~\cite{yan2002gspan}, can find subgraphs within a graph database that appear with a frequency greater than a predefined minimum support threshold.
However, setting a minimum support threshold is extremely difficult in security contexts.
If set too high, subtle but critical vulnerability patterns are missed, whereas setting it too low will cause the algorithm to find small and non-meaningful subgraphs \cite{jiang2024app, cheng2009identifying}. 
To address this, TKG~\cite{fournier-viger2019tkgd} has been proposed to dynamically and heuristically adjust the internal support threshold during mining, pruning the search space to efficiently return only the \textit{k} most frequent subgraphs.
We customized TKG~\cite{fournier2019tkg} with an \textit{unsafe root node constraint}, restricting all path extensions to edges adjacent to the designated \verb|LITERAL:UNSAFE_MODEL_NAME| node. 
This constraint ensures that every discovered subgraph contains at least one vulnerable model reference, filtering out other patterns that arise from the density of CPGs.
To use TKG with our dataset, we converted normalized graphs to the gSpan graph transaction format~\cite{yan2002gspan}, where each graph was represented as a sequence and normalized string labels were mapped to integer identifiers to reduce memory overhead.
TKG is governed by two parameters: \textit{k}, the number of frequent subgraphs to return, and \textit{max\_edges}, the maximum number of edges in a given subgraph.
\textrevision{Since the number of outputs increases exponentially with pattern size~\cite{yan2002gspan,fournier2019tkg,cheng2009identifying}, we explored \(k\in\{50, 100, 200\}\) and \(max\_edges\in[1, 50]\) for each framework to ensure tractability, while still capturing meaningful patterns beyond trivial relationships.
We observed that \(k=50, max\_edges=50\) fully captured the pattern search space: TKG saturated with no patterns exceeding 48 edges. 
Since TKG maintains a dynamically adjusted minimum support threshold corresponding to the \(k\)-th pattern, larger \(k\) introduced low-support, isomorphic subgraphs that were subsequently removed during post-processing. 
Thus, increasing \(k\) or \(max\_edges\) further did not surface additional patterns, and we adopted this configuration for subsequent analysis.}
Finally, we parsed the outputs and mapped integer identifiers back to their original labels for further post-processing.
For example, a mined pattern \texttt{LITERAL:UNSAFE\_MODEL\_NAME}-[\texttt{ARG}]$\rightarrow$\texttt{CALL:PIPELINE} represents a recurring subgraph in which the model identifier is passed directly as an argument to the \texttt{pipeline()} API across multiple repositories.

\textbf{Pattern Post-Processing.}
FSM/TKG yielded multiple subgraph instances.
While these patterns are useful as evidence of recurrence, they can be redundant with minor lexical variation and structural differences \cite{jiang2024app, cheng2009identifying}.
To facilitate interpretability and examine recurring usage behaviors instead of examining individual subgraph instances, we post-processed the raw patterns to generalize similar structures with subgraph isomorphism and grouped them into \textbf{motifs}~\cite{milo2002network} according to the semantic meaning of the dominant edge types.
We use the term \textbf{motif} in the program-analysis sense of a recurring and interpretable structural schema~\cite{narayanan2012effects}.
First, to resolve structural redundancy, we extracted the maximal patterns~\cite{thomas2010margin} similar to prior works~\cite{wi2022hiddencpg, hellal2015maximal, jiang2024app}, in which a pattern $P_i$ was removed if there existed another pattern $P_j$ ($i \neq j$) such that $P_i$ was isomorphic to a subgraph of $P_j$.
We employed NetworkX's VF2 implementation~\cite{networkx, cordella2004sub, foggia2001improved} for this step.
After maximal filtering, structurally equivalent patterns could differ only in their leaf node labels (e.g., \texttt{CALL:TRANSFORMERS} vs.\ \texttt{CALL:DIFFUSERS} at the same position).
To improve interpretability, we generalized each maximal pattern into an abstract template by replacing leaf nodes with wildcard labels while preserving internal nodes and edge types. 
This abstraction step conceptually aligned with prior work that generalized graph pattern summarization~\cite{dmitriev2006mining, cheng2010discriminative, yamaguchi2015pattern} to merge structurally similar instances into more compact and interpretable pattern classes.
In our setting, this step served to preserve core propagation structures and suppress minor variations.
Patterns were grouped into the same template if they shared the same abstract edge multiset after wildcard substitution.
We recorded the replaced leaves, the number of constituent patterns, and aggregated support statistics.

\textrevision{\textbf{Motif Taxonomy Construction.} 
To facilitate interpretability, we organized patterns into a motif taxonomy by classifying them based on the dominant semantic structure and manually assigning each resulting structure a descriptive name.
Since our CPG processing retained only the most semantically meaningful edge between nodes, each extracted pattern had one dominant semantic edge/node following the unsafe root. 
Patterns were automatically and deterministically classified into recurring motif structures based on these dominant edges/nodes, excluding those containing only \texttt{CFG}/\texttt{AST} edges, as these captured only execution ordering or syntactic co-location~\cite{yamaguchi2014modeling} and carried no semantic meaning relevant to usage motifs.
The first author then interpreted the semantics of each structure and proposed a name, which was then discussed and finalized collectively by all authors.
This resulted in the motif taxonomy presented in Section~\ref{subsec.res.rq2}.}

\section{Results}
\label{sec.res}

\subsection{\textrevision{Vulnerable Model Adoption and Propagation}}
\label{subsec.res.rq1}

Despite the availability of documentation on the dangers of vulnerable models from both the community~\cite{huggingface_pickle_scanning, python_pickle_module, protectai_insights, jfrog_model_threats} and academia~\cite{kellas2025pickleball, zhu2025my, xiao2018security, zhao2024models}, our empirical analysis shows that the adoption of these models remains concerning in practice.
We examine the spread of these models across four dimensions: direct code usage in GH repositories, distribution via software packages, propagation through fine-tuning chains on HF, and deployment in HF Spaces applications.
Table~\ref{tab:rq1_summary} summarizes the key statistics of the prevalence of vulnerable models across these dimensions.

\textbf{Usage in open-source repositories and packages.}
Among the 16,977 vulnerable models, 3,757 (22.1\%) appear at least once in a GH repository.
In total, we found 21,313 code occurrences spread across 5,040 unique repositories, indicating that many repositories reference more than one flagged model.
\texttt{hf-internal-testing/\-tiny-random-OnnxStableDiffusionPipeline} is the most referenced model with 976 occurrences in 350 repositories, but it is a test model from HF and is primarily used in CI/CD pipelines rather than in production deployments.
Other than the testing models, \texttt{lxyuan/distilbert-student}, illustrated in Figure~\ref{fig:propagation_example},
is the most widely used with 594~occurrences in 449~repositories.
While the HF scanner identified  as confirmed malicious artifacts, they remain deeply embedded in the developer ecosystem.
Beyond direct code usage, 68 flagged models (0.4\%) are referenced in repositories that produce software packages, yielding a total of 185 packages across 111 repositories.
These packages are predominantly container images distributed via GH Packages, suggesting that vulnerable models can propagate into containerized deployments and, consequently, into production environments.

\begin{table}[t]
\caption{Summary of vulnerable model prevalence across the AI supply chain (out of 16,977 flagged models).}
\resizebox{\columnwidth}{!}{%
\setlength{\tabcolsep}{3pt}
\begin{tabular}{p{1.7cm} l r r}
\hline
\multicolumn{2}{l}{\textbf{Supply Chain Dimensions}} & \textbf{Count} & \textbf{\%} \\ \hline

\multicolumn{4}{l}{\textit{GitHub Code Usage}} \\
\multicolumn{2}{l}{Vulnerable models with $\geq$1 code reference} & 3,757 & 22.1 \\
\multicolumn{2}{l}{Unique repositories with vulnerable models}    & 5,040 & --   \\
\multicolumn{2}{l}{Total code occurrences}                    & 21,313 & --   \\ \hline

\multicolumn{4}{l}{\textit{GitHub Package Distribution}} \\
\multicolumn{2}{l}{Vulnerable models in package-producing repos} & 68  & 0.4 \\
\multicolumn{2}{l}{Unique package-producing repos}           & 111 & -- \\
\multicolumn{2}{l}{Total unique packages}                    & 185 & -- \\ \hline

\multicolumn{4}{l}{\textit{Fine-Tuning Chains}} \\
\multicolumn{1}{l|}{\multirow{3}{*}{Upstream}}
& Vulnerable models with $\geq$1 parent & 3,765 & 22.2 \\
\multicolumn{1}{l|}{} & Total parent models               & 530   & --   \\
\multicolumn{1}{l|}{} & Parent also flagged               & 57    & --   \\ \cline{1-4} 

\multicolumn{1}{l|}{\multirow{4}{*}{Downstream}}
 & Vulnerable models with $\geq$1 child & 136 & 0.8 \\
\multicolumn{1}{l|}{} & Total child models               & 419 & --  \\
\multicolumn{1}{l|}{} & Children also flagged            & 43  & --  \\
\multicolumn{1}{l|}{} & Downloads across children        & 22,963 & -- \\ \hline

\multicolumn{4}{l}{\textit{HuggingFace Spaces}} \\
\multicolumn{2}{l}{Vulnerable models in $\geq$1 Space}     & 657  & 3.9 \\
\multicolumn{2}{l}{Total model--space usages}          & 1,766 & --  \\
\multicolumn{2}{l}{Unique Spaces with vulnerable models}   & 1,239 & -- \\
\multicolumn{2}{l}{Total Space likes}                 & 10,837 & -- \\
\end{tabular}
}
\label{tab:rq1_summary}
\end{table}

\textbf{Propagation through fine-tuning chains.}
On the upstream side, 3,765 (22.2\%) flagged models declare a parent or base model on HF, and 57 of these have a parent that is itself flagged, indicating that vulnerabilities can be inherited through the fine-tuning lineage.
On the downstream side, 136 flagged models (0.8\%) have been used as the base for at least one fine-tuned child model, producing a total of 419 downstream derivatives.
Among these, 43 children are also flagged, and the full set of child models has been downloaded 22,963 times collectively.
The most widely reused model is \texttt{Ultralytics/YOLO11} with 100 direct children and over 54,000 downloads.
Other notable cases include \texttt{IAmSkyDra/BARTBana\_v5} and \texttt{medalpaca/medalpaca-7b} with 50 and 32 fine-tuned variants, respectively.
\texttt{lxyuan}/\allowbreak\texttt{distilbert-student}, our illustrative example, is fine-tuned into 24 children with 11,925 total downloads.
This finding highlights that the impact of a single vulnerable model can be amplified considerably through fine-tuning chains.

\textbf{Deployment in HF Spaces.}
Finally, we examine whether vulnerable models are deployed in HF Spaces web applications that serve model inference to end users.
We find that 657 flagged models (3.9\%) are referenced in at least one Space, producing 1,766 model-space usage links across 1,239 unique Spaces.
These Spaces have accumulated 10,837 likes, indicating non-trivial user engagement.
The predominant SDKs used with these vulnerable models are Gradio (1,019 usages), followed by Docker (113), Streamlit (84), and static deployments (19).
This finding is concerning because Spaces represent the user-facing side of the supply chain where models interact with end users, potentially allowing the vulnerability to be directly exploited.

\textbf{Cross-layer Exposure.}
We observe that the four dimensions above are not isolated, and a single vulnerable model can propagate across all of them simultaneously, creating correlated risk that individual security views cannot properly capture.
We illustrate this with the motivating example introduced in Section~\ref{sec.related}, \texttt{lxyuan/distilbert-student} (Figure~\ref{fig:propagation_example}), as this is the non-test vulnerable model with the highest code-level adoption in our dataset.
The downstream footprint of this model spans all four dimensions at once.
It appears in 449 repositories, including well-established projects, e.g., \texttt{spark-nlp}~\cite{spark-nlp-github} from \texttt{JohnSnowLabs} and \texttt{activepieces}/\texttt{activepieces}~\cite{activepieces_repo}.
Specifically, the \texttt{activepieces} repository uses the vulnerable model directly in its HF integration plugin~\cite{activepieces_text_classification_action}, and this source code is then used to publish the container images via GH Packages.
These references confirm both the presence of vulnerable models in actively maintained systems and their propagation into production-level artifacts.
Furthermore, this model is also used in 82 HF Spaces and fine-tuned into 24 child models, where in some instances, the model is loaded directly~\cite{yuvarajareddy_youtube_sentiment_space}.
This can trigger pickle deserialization of the original PyTorch checkpoint on the server, a vector through which the \texttt{PYTCH-100} vulnerability of the model ~\cite{protectai2024distilbert-sentiment-overview} could potentially be exploited.
This represents a plausible risk pathway in which an unsafe model artifact may be deserialized at runtime on HF infrastructure and subsequently serve inference results to end users.
Moreover, among all 24 variants, only four have migrated to the \texttt{SafeTensor} format while others inherited the \texttt{.bin} format. 
Most variants were not scanned; however, several triggered pickle import warnings \cite{wnic00_distilbert_p1}, suggesting that the \texttt{PYTCH-100} vulnerability may have propagated further.
Collectively, the presence of vulnerable models across the ecosystem reveals a mismatch between model-level warnings issued by security advisories and ecosystem-level exposure. 
Warnings remain anchored to model artifacts while exposure propagates across multiple reuse surfaces without a unified provenance view~\cite{wang2024mitigating}.

\subsection{\textrevision{Integration Motifs and Security Implications}}
\label{subsec.res.rq2}

\begin{table*}[t]
{
\caption{Taxonomy of vulnerable model usage motifs.
For each motif, we present the defining graph patterns, the representative patterns and the maximum support for each framework.
Framework attribution is based on the project’s dominant framework.
In several cases, the sink API of the representative pattern belongs to a lower-level framework (e.g., HF Hub, Transformers).
Notes: \texttt{``*''} denotes an abstract placeholder in generalized patterns.
}
\label{tab:result_rq2}
\centering
\resizebox{\textwidth}{!}{%
\fontsize{8}{7}\selectfont
\begin{tabular}{@{}c@{}|l|l|c|@{}l}
\textbf{Motif} & \textbf{Framework} & \multicolumn{1}{c|}{\textbf{Representative CPG Slice}} & \textbf{\#} & \multicolumn{1}{c}{\textbf{Code Snippet}} \\  \hline
\multirow{12}{*}{\rotatebox[origin=c]{90}{Direct Usage (M1)}} &
\textit{Generalized} &
\textbf{\texttt{LITERAL:UNSAFE\_MODEL\_NAME} ${\xrightarrow{ARG}}$ \texttt{*}} 
& -- & \multirow{12}{*}{
\begin{minipage}[t]{0.25\linewidth}
\ttfamily \color{black!80}
{\fontsize{7}{9}\selectfont \#1: xzxg001/HNU-AI-Experiment} \\
\colorbox{gray!10}{%
\parbox{\linewidth}{%
model = AutoModelForSe\-quenceClassification.\linebreak from\_pretrained(
\textcolor{red}{"UNSAFE\_MODEL"},
trust\_remote\_code=\textcolor{red}{True})
}}
\\[4pt]
{\fontsize{7}{6}\selectfont \#2: ZCen-Xiong/GCL}
\colorbox{gray!10}{%
\parbox{\linewidth}{%
dataset = load\_dataset(\\\textcolor{red}{"UNSAFE\_MODEL"})
}} 
\\[4pt]
{\fontsize{7}{6}\selectfont \#3: Vince-Liuss/literary\_\linebreak style\_model}
\colorbox{gray!10}{%
\parbox{\linewidth}{%
style\_model = SentenceTransformer(\\
 \textcolor{red}{"UNSAFE\_MODEL"},\\
trust\_remote\_code=\textcolor{red}{True})
}}
\end{minipage}} \\
 & Transformer     & ${\xrightarrow{ARG}}$ \texttt{CALL:AUTOMODELFORSEQCLASS/AUTOTOKENIZER/PIPELINE} 
 & 121 & \\
 & HF Hub & ${\xrightarrow{ARG}}$ \texttt{CALL:HF\_HUB\_DOWNLOAD/SNAPSHOT\_DOWNLOAD} & 66 & \\
 & Ultralytics      & ${\xrightarrow{ARG}}$ \texttt{CALL:HF\_HUB\_DOWNLOAD/YOLO} 
 & 25 & \\
 & PyTorch          & ${\xrightarrow{ARG}}$ \texttt{CALL:LOAD} ${\xrightarrow{CDG}}$  \texttt{CONTROL\_STRUCTURE:IF} & 24 & \\
 & Diffusers        & ${\xrightarrow{ARG}}$ \texttt{CALL:HF\_HUB\_DOWNLOAD/SNAPSHOT\_DOWNLOAD} & 9  & \\
 & ONNX             & ${\xrightarrow{ARG}}$ \texttt{CALL:HF\_HUB\_DOWNLOAD/SNAPSHOT\_DOWNLOAD/LAYOUTMODEL/TAGGERMODELS} & 7 & \\
 & TensorFlow       & ${\xrightarrow{ARG}}$ \texttt{CALL:SNAPSHOT\_DOWNLOAD} & 5 &  \\
 & Keras            & ${\xrightarrow{ARG}}$ \texttt{CALL:HF\_HUB\_DOWNLOAD} ${\xrightarrow{RD}}$ \texttt{CALL:SAVING} & 5  & \\
 & Joblib           & ${\xrightarrow{ARG}}$ \texttt{CALL:HF\_HUB\_DOWNLOAD} & 3 & \\
 & Datasets         & ${\xrightarrow{ARG}}$ \texttt{CALL:ARGUMENTPARSER/LOAD\_DATASET} & 2 & \\
 & ST               & ${\xrightarrow{ARG}}$ \texttt{CALL:SENTENCETRANSFORMER} & 1 & \\
 \hline
\multirow{17}{*}{\rotatebox[origin=c]{90}{Local Propagation (M2)}} & \textit{Generalized} & 
\textbf{\texttt{LITERAL:UNSAFE\_MODEL\_NAME} ${\xrightarrow{RD}}$ \texttt{CALL:*}} & -- &
\multirow{17}{*}{
\begin{minipage}[t]{0.25\linewidth}
\ttfamily \color{black!80}
{\fontsize{7}{6}\selectfont \#4: d-senyaka/sinLlama} \\
\colorbox{gray!10}{%
\parbox{\linewidth}{%
BASE = \textcolor{red}{"UNSAFE\_MODEL"} \\[3pt]
ADAPTER = \textcolor{red}{"UNSAFE\_MODEL"} \\[3pt]
\textcolor{orange}{model} = AutoModelFor\-CausalLM.from\_pretrained( 
\textcolor{red}{BASE}, ... ) \\[3pt]
\textcolor{orange}{model} = PeftModel.from\_ \linebreak pretrained(\textcolor{orange}{model}, \textcolor{red}{ADAPTER}, ...)
}}
\\[4pt]
{\fontsize{7}{6}\selectfont \#5: Himanshu0518/Agroassist} \\
\colorbox{gray!10}{%
\parbox{\linewidth}{%
\textcolor{orange}{model\_path} = hf\_hub\_download(
repo\_id=\textcolor{red}{"UNSAFE\_MODEL"},)
\\[3pt]
joblib.load(\textcolor{orange}{model\_path})
}}

\end{minipage}} \\
 & Transformers & ${\xrightarrow{RD}}$ \texttt{CALL:AUTOMODELFORCAUSALLM/AUTOMODELFORSEQCLASS/AUTOTOKENIZER} & 115 \\
 & Sherpa       & ${\xrightarrow{RD}}$ \texttt{CALL:HF\_HUB\_DOWNLOAD/GET\_NN\_MODEL\_FILENAME/GET\_TOKEN\_FILENAME} & 76  \\
 & HF Hub & ${\xrightarrow{RD}}$ \texttt{CALL:HF\_HUB\_DOWNLOAD} & 60 \\
 & Ultralytics  & ${\xrightarrow{RD}}$ \texttt{CALL:YOLO} & 21 \\
 & ONNX         & ${\xrightarrow{RD}}$ \texttt{CALL:HF\_HUB\_DOWNLOAD/SNAPSHOT\_DOWNLOAD/UPSCALER\_MODELS} &  11 \\
 & PEFT         & ${\xrightarrow{RD}}$ \texttt{CALL:PEFTMODEL} & 9 \\
 & Diffusers    & ${\xrightarrow{RD}}$ \texttt{CALL:PATH} ${\xrightarrow{CDG}}$ \texttt{CONTROL\_STRUCTURE:IF} ${\xleftarrow{CDG}}$ \texttt{CALL:SNAPSHOT\_DOWNLOAD}   & 6 \\
 & TensorFlow   & ${\xrightarrow{RD}}$ \texttt{CALL:SNAPSHOT\_DOWNLOAD}     & 5  \\
 & Joblib       & ${\xrightarrow{RD}}$ \texttt{CALL:CREATE\_REPO/HF\_HUB\_DOWNLOAD/LOAD} & 4 \\
 & ST & ${\xrightarrow{RD}}$ \texttt{CALL:MODELS/SENTENCETRANSFORMER/SPARSETRANSFORMER} & 4 \\
 & Keras        & ${\xrightarrow{RD}}$ \texttt{CALL:HF\_HUB\_DOWNLOAD} ${\xrightarrow{RD}}$ \texttt{CALL:LOAD}  & 3\\
 & Optimum      & ${\xrightarrow{RD}}$ \texttt{CALL:AUTOTOKENIZER} & 3 \\
\hline
\multirow{4}{*}{\rotatebox[origin=c]{90}{\shortstack{Method\\Prop. (M3)}}} & \textit{Generalized} &
\textbf{\texttt{LITERAL:UNSAFE\_MODEL\_NAME} ${\xrightarrow{RD}}$ \texttt{METHOD:*}} & -- &
\multirow{4}{*}{
\begin{minipage}[t]{0.25\linewidth}
\ttfamily \color{black!80}
\vspace{-5pt}
{\fontsize{7}{6}\selectfont \#6: yih97/InsightSmith} \\
\colorbox{gray!10}{%
\parbox{\linewidth}{%
def yolo\_detector(path):

\hspace{5pt} path = hf\_hub\_download(

\hspace{5pt} \textcolor{red}{"UNSAFE\_MODEL"})

\hspace{5pt} model = YOLO(path)}}
\end{minipage}} \\
 & Ultralytics  & ${\xrightarrow{RD}}$ \texttt{METHOD:YOLO\_DETECTOR} ${\xrightarrow{CALL}}$ \texttt{CALL:YOLO} & 5 \\
 & Diffusers    & ${\xrightarrow{RD}}$ \texttt{METHOD:AD\_MODEL\_PROCESS} ${\xrightarrow{RD}}$ \texttt{CALL:MASK\_DILATE} & 3 \\
 &  &  & \\
\hline
\multirow{8}{*}{\rotatebox[origin=c]{90}{\shortstack{Condition-Guarded\\Propagation (M4)}}} & \textit{Generalized} &
\textbf{\texttt{LITERAL:UNSAFE\_MODEL\_NAME} ${\xrightarrow{CDG}}$ \texttt{*}} & -- &
\multirow{8}{*}{
\begin{minipage}[t]{0.25\linewidth}
\ttfamily \color{black!80}
{\fontsize{7}{6}\selectfont \#7: Yundi-Zhang/ViTa} \\
\colorbox{gray!10}{%
\parbox{\linewidth}{%
\textcolor{orange}{resume\_ckpt\_path} = download\_checkpoints(
repo\_id=\textcolor{red}{"UNSAFE\_MODEL"},) \\[3pt]
if not resume\_ckpt\_path:

\hspace{5pt} state\_dict = torch.load(

\hspace{10pt}\textcolor{orange}{resume\_ckpt\_path},

\hspace{10pt}weights\_only=\textcolor{red}{False})
}}

\end{minipage}} \\
 & Other        & ${\xrightarrow{CDG}}$ \texttt{CONTROL\_STRUCTURE:IF} & 34 \\
 & PyTorch      & ${\xrightarrow{CDG}}$ \texttt{CONTROL\_STRUCTURE:IF} ${\xrightarrow{CDG}}$ \texttt{CALL:LOAD} & 27 \\
 & HF Hub & ${\xrightarrow{CDG}}$ \texttt{CONTROL\_STRUCTURE:IF} ${\xrightarrow{CDG}}$ \texttt{CALL:SNAPSHOT\_DOWNLOAD}  & 19  \\
 & Diffusers    & ${\xrightarrow{CDG}}$ \texttt{CONTROL\_STRUCTURE:IF} ${\xrightarrow{CDG}}$ \texttt{CALL:SNAPSHOT\_DOWNLOAD} ${\xrightarrow{RD}}$ \texttt{CALL:PATH} & 6 \\
 & Ultralytics  & ${\xrightarrow{CDG}}$ \texttt{CONTROL\_STRUCTURE:IF} ${\xrightarrow{CDG}}$ \texttt{CALL:HF\_HUB\_DOWNLOAD} ${\xrightarrow{RD}}$ \texttt{CALL:YOLO} & 6 \\
\end{tabular}%
}}
\end{table*}

\textrevision{Our TKG mining step (Section~\ref{subsec.method.graph}) yielded 809 patterns.
After post-processing, 195 maximal patterns remained.
Our motif construction process allowed the patterns to be interpreted as a compact motif taxonomy based on the most semantically informative CPG relations~\cite{yamaguchi2014modeling, joern_cpg_spec}:}
\begin{itemize}[leftmargin=*]
\item \textbf{Direct Usage (M1)}: \texttt{ARGUMENT}-dominant patterns where the vulnerable model identifier is passed directly to an API, representing the shortest path from identifiers to sinks.
\item \textbf{Local Propagation (M2)}: \texttt{REACHING\_DEF}-dominant patterns, where the identifier is first staged as a local value and then propagated within the same scope.
\item \textbf{Method Propagation (M3)}: inter-procedural propagation patterns in which \texttt{RD} edges cross a \texttt{METHOD} boundary before the identifier reaches a downstream sink.
\item \textbf{Condition-Guarded Propagation (M4)}: \texttt{CDG}-dominant patterns, where the use of vulnerable model identifiers is primarily governed by branch control logic.
\end{itemize}

\noindent \textrevision{This taxonomy is illustrated by Table~\ref{tab:result_rq2}, with patterns and classification scripts available in our replication package~\cite{replication_package_vul_model_usage}.}
Following, we present the motifs, the security risks they implicate, and the real-world scenarios in which they appear.

\subsubsection{\textbf{Motif 1 (M1): Direct Usage}}
M1 captures patterns where vulnerable model names are passed directly as arguments to sink APIs, with no intermediate staging or wrapper boundary.
Although some patterns include additional controls or structural contexts, the defining structure are \texttt{ARGUMENT} edges between \texttt{LITERAL:UNSAFE\_MODEL\_NAME} and sinks, representing the shortest possible source-to-sink path. 
As shown in Table~\ref{tab:result_rq2}, M1 appears across 11 frameworks, with the highest support in Transformers and HF Hub (121 and 66 occurrences).
Destination sinks surface as two main recurring roles: \textit{artifact acquisition}, e.g., \texttt{hf\_hub\_download}, \texttt{snapshot\_download}, and \texttt{load\_dataset}, and \textit{artifact consumption}, e.g., \texttt{pipeline}, \texttt{load\_model}, and \texttt{YOLO}. 
This coverage suggests that direct sink binding is not limited to a single framework or API family, but is instead a common integration pattern across ML ecosystems.

\textbf{Security Implication:} The main security relevance of this motif lies in minimal source-to-sink distance, as there is no local variable, method boundary or conditional guard visible between the vulnerable identifier and the destination call. 
Any security configuration is delegated to call-site parameters or framework defaults.
In frameworks where validation is optional or parameter-dependent, this usage creates the possibility that safety checks remain unenforced.

\textbf{Examples:} Several of our collected repositories consist of this motif.
A representative security-relevant example is presented in Table~\ref{tab:result_rq2}, Snippet 1, where the vulnerable model name is passed directly to \texttt{from\_pretrained} together with \texttt{trust\_remote\_code=True}.
The Transformers documentation describes \texttt{trust\_remote\_code} as a parameter that permits execution of arbitrary code during model loading from remote repositories. 
The presence of such flag at the same call site with the vulnerable identifier demonstrates the decision of the developer to explicitly allow remote code execution with no visible intermediate step.
In \texttt{SentenceTransformers} (Table~\ref{tab:result_rq2}, Snippet 3), the same co-occurrence appears where \texttt{SentenceTransformer} receives both the flagged identifier and the trust flag at instantiation.
In the \textit{Datasets} framework, the identifier is also passed directly into the dataset loading (Table~\ref{tab:result_rq2}, Snippet~2), resolving the vulnerable model name immediately with no local validation logic.
By utilizing direct parameter usage without further configuration, developers completely omit the \texttt{verification\_mode} parameter. 
The Datasets documentation~\cite{huggingface_datasets_loading_methods} explicitly notes that \texttt{verification\_mode} checks for downloaded dataset metadata, including checksums, size, and split information. 
This parameter
defaults to a rudimentary check on remote metadata and skips downloaded file integrity (checksums, correct number of files downloaded) in favor of downloading performance.
As a result, such usage leaves stronger integrity verification on framework defaults rather than making it explicit at the call site.
These examples illustrate cases where vulnerable models reach sink APIs without any safeguards, and security or integrity implementations are delegated to destination frameworks.

\subsubsection{\textbf{Motif 2 (M2): Local propagation}}
This motif captures patterns where the vulnerable model identifiers are first materialized as local program states and then propagated through one or more data-flow steps before reaching a sink.
Similar to M1, some patterns consist of conditional checks but do not explicitly aim to address the integrity or validate the vulnerable models. 
The core structure of M2 is a \texttt{REACHING\_DEF} edge connecting the vulnerable identifiers to the call sites.
Thus, these patterns introduce source-to-sink distance through aliasing, local staging, path assembly, or configuration-mediated flow.
In our dataset, M2 is the most broadly distributed (12 frameworks) with highest support in Transformers~(115), Sherpa~(76), and HF Hub (60).

\textbf{Security Implication:} 
The main security concern of M2 is that vulnerability-relevant identifiers become obscured within local program state through aliasing, reassignment, or intermediate variables.
Once the identifier is staged as an ordinary variable, downstream sink usages may appear disconnected from the original unsafe source unless the local data-flow chain (intra-procedural) is reconstructed.
This can hinder static analysis tools and manual inspection by requiring reconstruction of local variable-level propagation within the same scope, even when no function boundary is crossed.
Thus, M2 highlights how vulnerable model identifiers become absorbed into ordinary program states.

\textbf{Examples:} In our dataset, M2 most commonly appears in three sub-patterns.
The first is \textit{alias-and-forward chains}, where the identifier is bound to a local variable and forwarded into one or more loading calls. 
An example appears in PEFT-style model composition (Table \ref{tab:result_rq2}, Snippet~4), where both a base model identifier and an adapter identifier are bound to local variables (\texttt{BASE/ADAPTER}) before being passed into successive \texttt{from\_pretrained} calls in \texttt{AutoModel} or \texttt{PeftModel}.
The flagged identifier is not consumed immediately but staged as a variable and then propagated into multiple loading operations across both the base model and adapter path.
This distributes the security configurations across a short local flow chain rather than concentrating it at a single literal-callsite binding like M1.
The second sub-pattern is \textit{local path construction}, where the identifier determines a downloaded artifact path that is then forwarded into a loading call.
An example appears in a Joblib-style loading workflow (Table \ref{tab:result_rq2}, Snippet 5), where the flagged identifier is passed to \texttt{hf\_hub\_download} as \texttt{repo\_id}, and the resulting \texttt{model\_path} is then forwarded into \texttt{joblib.load}.
The vulnerable identifier does not appear directly at the sink, but rather routed through the local path variable.
\texttt{joblib.load}, the terminal sink in this propagation chain, also raises a security concern, as the documentation \cite{joblib_load} explicitly warns that it relies on the pickle module and should never be used to load files from untrusted sources and has also been discouraged by the literature \cite{casey2024large, kellas2025pickleball}.
A similar pattern appears in the \texttt{keras} framework, where the model is also downloaded via HF
and forwarded to a loading call.
The Keras documentation \cite{keras_model_saving_loading} notes that \texttt{load\_model} exposes a \texttt{safe\_mode} parameter which, when set to \textit{False}, permits arbitrary code execution during deserialization.
In all Keras M2 patterns, this parameter is absent from the call site, suggesting that security configurations are still being delegated to the framework defaults rather than made explicit by the developer.
The third sub-pattern is \textit{configuration} or \textit{parser routing}, most visible in Sherpa (76 occurrences), where the identifier is staged through model filename retrieval calls (\texttt{get\_nn\_model\_filename, get\_token\_filename}) before reaching the actual loading.
This is the most opaque sub-pattern in M2, where the identifier is absorbed into framework-internal configuration structures before reaching any sink.
Across all three sub-patterns, vulnerable model names become ordinary local values before reaching security-sensitive APIs, reducing the visibility of their provenance at the sink.

\subsubsection{\textbf{Motif 3 (M3): Method propagation}}
M3 captures patterns where vulnerable model identifiers cross at least one application-defined method before reaching sink APIs.
The identifier is routed through custom wrappers, helper functions, or builder methods, abstracting sink usages with application-specific logic rather than explicit call sites.

\textbf{Security Implication:} 
Compared to M1/M2, this motif further reduce immediate sink visibility and distribute security logic across multiple abstraction layers.
However, unlike local propagation (M2), the data-flow is now distributed across multiple call sites (inter-procedural) since they are routed through multiple methods.
As a result, security analysis must reconstruct the complete inter-procedural source-to-sink paths to accurately identify the vulnerability propagation.

\textbf{Examples:} In our dataset, M3 appears mainly in Ultralytics and Diffusers with 5 and 3 occurrences, respectively.
Both frameworks exhibit the same high-level structure: unsafe models-[RD]$\rightarrow$\texttt{METHOD:*}-[RD/CALL]$\rightarrow$\texttt{CALL:*}, but differ in the nature of wrappers and terminal sinks.
In the Ultralytics pattern example (Table~\ref{tab:result_rq2}, Snippet~6), the flagged identifier propagates into a user-defined \texttt{yolo\_detector} wrapper method, which internally calls \texttt{hf\_hub\_download} to retrieve the artifact and then passes the resulting path to \texttt{YOLO} for instantiation.
The identifier is not visible at the final \texttt{YOLO} call, but rather consumed inside the wrapper, and the only artifact that reaches the terminal sink is the downloaded path.
This wrapper structure means that the propagation chain traverses a method boundary before reaching the HF Hub download layer and then a second implicit step into the \texttt{YOLO} loader.
However, even with user-defined wrapper functions, security configurations are still being delegated to internal framework mechanisms, which is especially concerning given the recent supply chain attack on YOLO~\cite{ultralytic_attack, snyk_ultralytics, hiddenlayer_ultralytics}.
The \texttt{YOLO} constructor internally invokes PyTorch's model loading, which relies on pickle deserialization~\cite{ultralytics_model_py, kellas2025pickleball}.
The attack specifically targeted this deserialization logic in \texttt{model.py} and \texttt{downloads.py}~\cite{hiddenlayer_ultralytics}, the same files that implement the artifact retrieval steps that this pattern showcases, to run crypto-mining malware before model loading. 
In Diffusers, the identifier propagates into \texttt{METHOD:AD\_MODEL\_PROCESS}, a model-processing wrapper, which then routes to \texttt{CALL:MASK\_DILATE}.
The inter-procedural hop separates the model origin from downstream calls, showing how wrapper methods load vulnerable models and expose only their outputs to subsequent call sites without visibility of the original identifier.

\subsubsection{\textbf{Motif 4 (M4): Condition-Guarded Propagation}}
M4 captures patterns in which the introduction or use of vulnerable model identifiers is governed primarily by branch control contexts.
The defining structural property is the \texttt{CDG} edge from the unsafe \texttt{LITERAL} to a \texttt{CONTROL\_STRUCTURE} node, meaning the identifier's loading or retrieval is conditional on a branch predicate.
M4 appears across five frameworks: Other (34 occurrences), PyTorch (27), HF Hub (19), Diffusers (6), and Ultralytics (6).
The high support in the Other category (34) is notable, as it suggests that condition-guarded loading is particularly common in user-defined code that does not resolve to a dominant framework.

\textbf{Security Implication:} 
Unlike M1-M3, which emphasize direct binding or propagation paths, M4 highlights contexts in which retrieval/loading behaviors are selected or triggered.
However, in all patterns, we observe that conditional guards reflect operational dispatch, e.g., selecting which model to load, checking whether a cached version exists, or gating download on a configuration flag.
We found no occurrence of explicit identity or integrity validation, e.g., cryptographic provenance checking, hash verification, or sanitization logic.

\textbf{Examples:} 
A representative example is documented in a \textit{PyTorch} checkpoint loading workflow (Table~\ref{tab:result_rq2}, Snippet~7).
In this example, the identifier determines a downloaded checkpoint path, and a subsequent conditional block passes that path to \texttt{torch.load} with \texttt{weights\_only=False}.
According to the official PyTorch documentation~\cite{pytorch_torch_load_doc, pytorch_weights_only_security}, \texttt{torch.load} uses an unpickler internally, and \texttt{weights\_only=False} explicitly enables the loading of arbitrary source code from models.
This pattern of a flagged deserialization-class model identifier with this parameter in a \texttt{CDG}-dominated pattern illustrates that the conditional structure does not validate the model's identity before the load call, and the parameter configuration at the load site explicitly disables the primary available safeguard.
Whether this constitutes an actual exploitable path depends on the runtime execution and the specific artifact loaded, but remains a security concern nonetheless.
In HF Hub patterns (19 occurrences), the chain extends to -[CDG]$\rightarrow$\texttt{CONTROL\_STRUCTURE:IF}-[CDG]$\rightarrow$\texttt{CALL:SNAPSHOT\_DOWNLOAD}, indicating that the condition only guards the artifact retrieval step. 
In Diffusers and Ultralytics (6 occurrences for both), the chains are similarly structured but extend to \texttt{CALL:SNAPSHOT\_DOWNLOAD}-[RD]$\rightarrow$\texttt{CALL:PATH} and \texttt{CALL:HF\_HUB\_DOWNLOAD}-[RD]$\rightarrow$\texttt{CALL:YOLO} respectively, showing conditioned downloads flowing directly into subsequent loading calls in the same branch.
The structural observation across M4 is consistent, where the presence of conditional guards do not constitute provenance or security checks.
These conditions only determine whether or which model to load given a runtime state, not whether to validate the model's identity or to guard the model loading process.

\subsubsection{\textbf{Cross-framework dependency}}
Although the motifs reveal distinct usage and propagation structures, Table~\ref{tab:result_rq2} also highlights a cross-cutting pattern: multiple frameworks rely on a shared set of lower-level retrieval and loading APIs.
In particular, Keras, TensorFlow, Ultralytics, Diffusers, ONNX, and Joblib recurrently expose patterns involving \texttt{hf\_hub\_download} or other HF loading routines, regardless of high-level framework APIs through which model identifiers are introduced.
At the application level, projects appear to use various APIs, e.g., \texttt{load\_model()}, \texttt{YOLO()}, or other framework-specific wrappers.
However, representative CPG slices show multiple usages either directly invoke HF Hub download functions or propagate into them through short local data-flow chains.
As a result, configurations that appear framework-specific often collapse onto a smaller set of shared sink APIs.
From a security perspective, this creates a form of cross-framework concentration: risks associated with identifiers of vulnerable models are not isolated within individual frameworks but instead converge on shared retrieval and loading APIs.
Despite this concentration, practical mechanisms for detecting unsafe uses of such identifiers remain limited, particularly in identifying security-relevant retrieval and loading APIs and tracing the data flow from unsafe model references into them, which further motivates data flow-aware analysis.
Conversely, this concentration may offer defensive leverage, as hardening a small number of shared sink APIs could yield benefits across multiple frameworks.
This convergence also strengthens the methodological value of the mined motifs, as recurring dependence on shared sink APIs suggests that FSM can help identify high-value and high-frequency sinks that are relevant ecosystem-wide.

\textrevision{\subsubsection{\textbf{Vulnerability-relevant Usage Analysis}}
To assess whether model references correspond to actual vulnerability-relevant usages, we analyzed model-usage patterns against vulnerabilities associated with 289 models in our RQ2 corpus: 237 deserialization, 59 runtime, and no backdoor threats.
We identified frameworks with loading paths applicable to these threats and analyzed the three most represented for deserialization (Transformers, PyTorch, and Ultralytics) and the two runtime-applicable frameworks (Keras and TensorFlow).
We considered a reference usage vulnerable when (1) the model was flagged by InsightsDB for a specific vulnerability, and (2) the vulnerable model were then loaded through a framework- or format-compatible path that is susceptible to the flagged vulnerability, and lacks the corresponding safeguards, or explicitly enables unsafe behaviors.
For example, a Pickle-deserialization vulnerability in a model is only applicable when the flagged Pickle artifact is consumed through an unsafe Pickle-compatible loader (e.g., \texttt{torch.load} without \texttt{weights\_only}).
Artifact-acquisition frameworks and their APIs, e.g, \texttt{huggingFace\_hub}, which only download model artifacts without consuming them, were excluded.
This analysis identified 554 vulnerability-relevant load paths, of which \textbf{553 (99\%)} matched the vulnerability conditions associated with referenced models.
We identified 533 vulnerable deserialization paths, with 272 using APIs without deserialization safeguards (\texttt{transformers.pipeline} and \texttt{ultralytics.YOLO}), while 261 omitted or disabled available safeguards, (\texttt{transformers.from\_pretrained} without \texttt{use\_safetensors} and \texttt{torch.load} without \texttt{weights\_only}). 
Separately, 48 Transformers paths explicitly enabled \texttt{trust\_remote\_code=True}, permitting remote code execution. 
Among 20 vulnerable runtime paths, 16 Keras usages invoked \texttt{load\_model/load\_weights} without \texttt{safe\_mode}, while four TensorFlow \texttt{load\_model} directly loaded unsafe \texttt{SavedModel} file format.}

\section{Discussion}
\label{sec.discussion}

\subsection{Security Provenance Gaps and Implications}

\subsubsection{Ecosystem-level provenance fragmentation}

Our RQ1 results show that the current ecosystem lacks mechanisms to connect model artifacts to their downstream exposure surfaces, limiting cross-layer visibility in the AI supply chain.
The same vulnerable models can persist across multiple layers of the AI supply chain simultaneously, i.e., code references, packaged artifacts, derivative models, and deployed applications, yet these reuse paths are only partially linked by existing metadata or tooling.
This gap aligns with prior calls for machine-readable provenance and ecosystem-level dependency graphs in open-source ecosystems.
Bommasani et al.~\cite{bommasani2024ecosystem} argue that such graphs are needed to support accountability, while Wang et al.~\cite{wang2024mitigating} show that existing provenance mechanisms do not provide a unified view linking advisory signals to downstream consumption.
Our results extend these limitations to the AI supply chain by showing that vulnerable model propagation paths only become visible when aggregating multiple unconnected data sources. 
Moreover, the propagation dimensions we observe correspond to distinct reuse surfaces for a single dependency, each governed by different tooling and visibility constraints.

While the results of RQ1 are comprehensive, our measured exposures are likely a lower bound.
Jiang et al.~\cite{jiang2024peatmoss} and Stager et al.~\cite{stalnaker2025empirical} show that artifact reuse is only partially documented across open-source ecosystems, with consumption occurring through configuration files, external downloads, and weak metadata that evade code-level detection.
The adoption and propagation risks we observe therefore parallel findings from traditional package ecosystems such as \texttt{npm} and \texttt{Maven}~\cite{decan2018impact, zimmermann2019small}, where risks propagate through transitive adoption with limited downstream visibility.
Our results extend this observation to the AI model supply chain, where propagation occurs through fine-tuning chains and deployed applications rather than explicit dependency declarations, amplifying risks without adequate lineage tracking.

\textrevision{Despite the availability of security advisories, vulnerable models are still being widely adopted in mature open-source workflows.
Developers may prioritize these models for performance or ease of use over security, or are insufficiently informed about security risks during model selection, similar to how vulnerable dependencies continue to be used for their convenience and familiarity in software ecosystems~\cite{kula2018developers, jafari2023dependency}.
Examining these behaviors, e.g., through developer surveys, is a useful direction for future work to help clarify security considerations during model selection.}
These challenges highlight that advisory availability alone is insufficient to reduce downstream exposure, as warnings can identify individual risky models but do not provide the cross-layer visibility needed to assess how far those risks have propagated.
Reducing exposure therefore requires automated downstream tracing and stronger provenance support across integration surfaces, rather than relying on model-level warnings alone.
A practical path forward is to complement provenance artifacts, e.g., AIBOM~\cite{nist2024securing}, with machine-readable vulnerability records for AI models, analogous to traditional software vulnerability disclosures, so that downstream hubs and tools can propagate inherited risk signals across fine-tuning, packaging, and deployment chains.
However, unlike traditional software components, such representations need to capture not only artifact-level vulnerabilities but also relationships across fine-tuning chains, derivative models, and integration contexts.

\subsubsection{Lack of security and provenance tooling during integration}

While the motifs in Section \ref{subsec.res.rq2} may represent common coding paradigms, they expose a recurring gap between model loading, validation, and security enforcement.
Across our motifs, vulnerable models reach retrieval or loading APIs through structures without explicit security decisions at the point of use.
This does not imply that every instance is exploitable or validation is entirely absent, but rather shows that security enforcements are often displaced into resolving frameworks. 
Collectively, these motifs reveal a pattern we characterize as \textit{provenance-blind integration}, where developers produce syntactically correct and idiomatic code that propagates potentially vulnerable models into execution contexts without local identity or integrity verification.
This pattern differs from software vulnerabilities, e.g., deserialization flaws, that arise from explicit coding errors.
Instead, it reflects the absence of security-enforcement steps that current APIs and documentation often do not mandate.

The frequency and consistency of the motifs suggest that the lack of trust-enforcement steps is not an individual coding choice, but a recurring outcome of how security enforcement is handled during integration.
While recent work has shown that model loading protections in ML frameworks are unevenly enforced~\cite{digregorio2026security}, our findings show how this fragmentation manifests during model integration in practice.
Specifically, there is a responsibility mismatch between frameworks, model hubs, and users.
Frameworks provide security controls that are often optional or not visible locally at the point where models are loaded.
For example, the \texttt{verification\_mode} in the Datasets framework enforces strong file verifications, but defaults to \texttt{BASIC\_CHECKS} which omits validation of downloaded files, e.g., checksum checks~\cite{huggingface_datasets_loading_methods}. 
HF documentation recommends that developers use trusted sources, review repository code, pin \texttt{revision} to a specific commit hash, and use \texttt{dry\_run} to inspect the metadata of the models being downloaded, but these configurations are also not explicitly enforced~\cite{huggingface_hf_hub_download}.
Platform advisories, e.g., ProtectAI, flag vulnerable models, yet those signals do not propagate into the code integration layer where loading decisions are made.
The application source code, in turn, integrates models through patterns that pass vulnerable identifiers to sinks without explicit local verification steps.

Such observations are in contrast to how analogous supply chain risks are handled in software ecosystems.
In npm or PyPI environments, audit tools, e.g., \texttt{pip-audit}, \texttt{npm audit}, Dependabot, and GH's dependency graph create feedback loops between security advisories and developer actions, backed by lockfile hash pinning, SBOM-style manifests, and IDE-level warnings at the point of import.
Equivalent mechanisms either do not exist or have not yet matured for AI model integrations.
There is no mechanism to bind a model identifier to a verified hash, no signal triggered when known vulnerable models are loaded, and no framework-level behaviors that block or warn against the loading of these vulnerable models.
Model cards~\cite{huggingface_model_cards, mitchell2019model} and AIBOM~\cite{nist2024securing} are the first steps towards improving transparency, yet they remain in infancy for capturing security and provenance information~{\cite{xia2024trust,radanliev2024capability,stalnaker2025understanding}}.

From a mitigation perspective, the motifs indicate that a single defensive strategy is likely insufficient, consistent with prior work on graph-based vulnerability discovery and API misuse mining~\cite{yamaguchi2014modeling, yamaguchi2015pattern, jiang2024app}.
M1 suggests stronger sink-level defaults and clearer high-risk API signaling.
M2 motivates data-flow-aware analysis beyond call-site matching.
M3 requires interprocedural tracing to recover hidden sinks.
M4 shows that branch logic should not be mistaken for trust enforcement.
These motifs therefore not only summarize how vulnerable models appear in code, but also identify distinct recurring controls where future mitigations can be applied.

\subsection{Recommendations}
\textbf{Security research application:}
\textrevision{Our findings suggest that FSM can complement CPG-based security workflows that typically rely on handcrafted traversals~\cite{wi2022hiddencpg,yamaguchi2014modeling}.
Manual queries require candidate sinks to be specified beforehand and may therefore overlook infrequent or previously unknown security-relevant APIs.
In RQ2, extracted CPGs contain an average of 2,793 nodes, 20,134 edges, with 3,055 distinct normalized node labels, while the most frequent labels are largely generic (e.g., \texttt{CONTROL\_STRUCTURE} and \texttt{LITERAL}), providing limited guidance for sink discovery.
FSM instead surfaces recurring propagation patterns that can guide subsequent CPG queries without exhaustive sink enumeration.
In our setting, the resulting motifs identify both recurring propagation structures and retrieval/loading APIs that act as candidate sinks across frameworks: M1 and M2 capture direct and short-range propagation, M3 captures inter-procedural propagation, and M4 captures branch-controlled usage contexts.
Table~\ref{tab:result_rq2} further shows how such motifs can expose cross-cutting sinks that would otherwise require multiple framework-specific queries and subsequent aggregation.
Moreover, FSM and VF2 account for only 30\% of our end-to-end runtime of 20 hours, indicating that this pattern-discovery stage is computationally feasible.}

\textrevision{To operationalize these motifs for security analysis, context-aware analysis can be utilized to model security-relevant behavior across frameworks and execution contexts.
For a model flagged with a Pickle-based deserialization vulnerability, a context-aware analyzer could (1) use framework semantics to recognize Transformers/PyTorch/Keras load paths that consume Pickle-format artifacts, (2) distinguish vulnerable load paths from unreachable ones, and (3) apply security policies to check whether safeguards, e.g., \texttt{use\_safetensors=True}, or \texttt{weights\_only=True}, should be enforced.
This would let usages be classified by contexts, reducing false positives while flagging configurations that warrant remediation, which is particularly important given practitioners’ concerns regarding false positives, missing context, and trust in AI-powered security tools~\cite{kholoosi2026software}. 
The cross-framework and framework-specific patterns in Table~\ref{tab:result_rq2} further motivate capturing these shared security primitives and framework-specific contexts.
However, adoption may be hindered by the maintenance effort, the diversity and rapid development of AI frameworks, and the absence of machine-readable and model-specific security advisories.}

\textbf{Practical application:}
Our findings suggest actionable directions for securing model integration pipelines.
First, security-sensitive loading settings should be made more explicit at the point of use.
Across motifs, security-related behavior is often left to framework defaults or optional flags, suggesting that high-risk APIs should require clearer opt-in decisions, stronger secure-by-default behavior, or explicit acknowledgment when permissive settings such as \texttt{trust\_remote\_code=True} or \texttt{weights\_only=False} are used.
Second, advisory and scanner signals should surface closer to the integration layer rather than isolated on model-hosting platforms.
For example, when a flagged model is requested for download through shared retrieval APIs, e.g., \texttt{hf\_hub\_download()}, users should receive explicit warnings, policy gates, or approval steps before execution proceeds to ensure proper acknowledgment.
Third, as multiple frameworks converge on a relatively small set of shared sink APIs, hardening these shared loading paths may provide ecosystem-wide defensive value.
Fourth, provenance artifacts, i.e., AIBOM, could be complemented with machine-readable vulnerability records for AI models to provide better visibility of ecosystem-level security, analogous to how VEX~\cite{cisa_vulnerability_exploitability} extends SBOM with structured vulnerability information.
Finally, our discovered motifs are usable in practice as the basis for developing static analysis rules, CI checks, IDE warnings, or automated remediation support~\cite{ye2025llmsecconfig} that identify or mitigate vulnerable model integrations before deployment~\cite{arani2023sok,arani2024systematic}.

\subsection{Limitations}
Our prevalence analysis relies on identifying code-level references of vulnerable model identifiers, rather than dynamic runtime analysis.
While these references allow us to locate vulnerable usage contexts, they do not guarantee runtime loading and execution.
Indirect usages, e.g., dynamically constructed identifiers, are also not captured, similar to incomplete vulnerability observations reported in prior empirical studies~\cite{le2024latent}.
Ecosystem-scale dynamic analysis requires executing untrusted code across diverse environments, which is computationally expensive, difficult to standardize, and poses safety risks.
Therefore, our prevalence analysis should be interpreted as an approximation of potential exposure surfaces, where the presence of vulnerable models may introduce risk to dependent systems, but does not imply actual or confirmed exploitation, consistent with prior evidence that security-analysis results can differ substantially under more realistic operational settings~\cite{nguyen2025toward}.

Our study focuses on models flagged by built-in security HF scanners and their propagation across GH and HF.
This scope excludes models from other platforms (e.g., PyTorch Hub) and threats not covered by the scanner, e.g., data poisoning.
While HF scanner does not capture all possible risks, it represents the ecosystem's primary automated security mechanism and serves as a practical, standardized reference to identify potentially vulnerable models.
Thus, our findings are constrained by scanning coverage and the reliability of automatically generated security labels, which may not reflect all relevant risks~\cite{le2024automatic}.

Finally, we cannot guarantee that discovered patterns generalize to all AI model integration scenarios.
Projects using proprietary frameworks, private model registries, or languages other than Python may exhibit different patterns.
However, since our dataset covers the dominant publicly available ML ecosystem (HF/GH) and includes diverse frameworks and real-world applications, it provides a representative baseline for understanding vulnerable model integrations.

\section{Conclusion}
\label{sec.conclusion}
We presented an empirical investigation of the usage and propagation of vulnerable models within the AI supply chain.
By analyzing the downstream usage of 16,977 vulnerable models, we showed that they are deeply embedded in the ecosystem with 21,000 references in more than 5,000 repositories, and further propagate through software packages, fine-tuning chains, and deployed applications.
To understand the mechanics of this exposure, we designed a custom CPG and FSM approach to extract recurring coding patterns that developers use when integrating these models.
We identified four primary usage motifs: Direct Usage, Local Propagation, Method Propagation, and Condition-Guarded Propagation, characterizing how vulnerable models are integrated.
Our findings reveal a concerning reality where, despite the availability of security advisories and framework-level safeguards, vulnerable models are frequently integrated without cryptographic provenance checking, hash verification, or sanitization logic.
The widespread tendency to delegate security to framework defaults, or to explicitly disable safeguards, highlights a critical gap in current AI development practices.
Our study underscores that artifact-level scanning alone is insufficient to secure the AI supply chain.
Addressing these threats require shifting security interventions closer to integration, developing context-aware static analysis tools capable of tracing complex data flows, and enforcing secure-by-default behaviors across ML frameworks.

\bibliographystyle{IEEEtran}
\bibliography{ref}

@inproceedings{ding2025rusty,
  title = {A Rusty Link in the {{AI}} Supply Chain: Detecting Evil Configurations in Model Repositories},
  booktitle = {2025 {{IEEE Security}} and {{Privacy Workshops}} ({{SPW}})},
  author = {Ding, Ziqi and Fu, Qian and Ding, Junchen and Deng, Gelei and Liu, Yi and Li, Yuekang},
  year = 2025,
  pages = {260--264},
  publisher = {IEEE},
  doi = {10.1109/SPW67851.2025.00036}
}

@misc{drewt2025pytorch,
  title = {{{PyTorch}} Users at Risk: Unveiling 3 Zero-Day {{PickleScan}} Vulnerabilities},
  author = {Drewt},
  year = 2025,
  journal = {JFrog},
  url = {https://jfrog.com/blog/unveiling-3-zero-day-vulnerabilities-in-picklescan/}
}

@incollection{fournier-viger2019tkgd,
  title = {{{TKG}}: Efficient Mining of Top-{{K}} Frequent Subgraphs},
  booktitle = {Big Data Analytics},
  author = {{Fournier-Viger}, Philippe and Cheng, Chao and Lin, Jerry Chun-Wei and Yun, Unil and Kiran, R. Uday},
  editor = {Madria, Sanjay and {Fournier-Viger}, Philippe and Chaudhary, Sanjay and Reddy, P. Krishna},
  year = 2019,
  volume = {11932},
  pages = {209--226},
  publisher = {Springer International Publishing},
  doi = {10.1007/978-3-030-37188-3_13}
}

@inproceedings{hu2025large,
  title={Large language model supply chain: Open problems from the security perspective},
  author={Hu, Qiang and Xie, Xiaofei and Chen, Sen and Quan, Lili and Ma, Lei},
  booktitle={Proceedings of the 34th ACM SIGSOFT International Symposium on Software Testing and Analysis},
  pages={169--173},
  year={2025}
}

@article{huang2025lifting,
author = {Huang, Kaifeng and Chen, Bihuan and Lu, You and Wu, Susheng and Wang, Dingji and Huang, Yiheng and Jiang, Haowen and Zhou, Zhuotong and Cao, Junming and Peng, Xin},
title = {Lifting the Veil on Composition, Risks, and Mitigations of the Large Language Model Supply Chain},
year = {2026},
journal={ACM Transactions on Software Engineering and Methodology},
}

@inproceedings{jiang2022empirical,
  title={An empirical study of artifacts and security risks in the pre-trained model supply chain},
  author={Jiang, Wenxin and Synovic, Nicholas and Sethi, Rohan and Indarapu, Aryan and Hyatt, Matt and Schorlemmer, Taylor R and Thiruvathukal, George K and Davis, James C},
  booktitle={Proceedings of the 2022 ACM workshop on software supply chain offensive research and ecosystem defenses},
  pages={105--114},
  year={2022}
}

@inproceedings{jones2024what,
  title={What do we know about Hugging Face? A systematic literature review and quantitative validation of qualitative claims},
  author={Jones, Jason and Jiang, Wenxin and Synovic, Nicholas and Thiruvathukal, George and Davis, James},
  booktitle={Proceedings of the 18th ACM/IEEE international symposium on empirical software engineering and measurement},
  pages={13--24},
  year={2024}
}

@inproceedings{li2021deeppayload,
  title={Deeppayload: Black-box backdoor attack on deep learning models through neural payload injection},
  author={Li, Yuanchun and Hua, Jiayi and Wang, Haoyu and Chen, Chunyang and Liu, Yunxin},
  booktitle={2021 IEEE/ACM 43rd International Conference on Software Engineering (ICSE)},
  pages={263--274},
  year={2021},
  organization={IEEE}
}

@inproceedings{li2025exploring,
  title={Exploring Typo Squatting Threats in the Hugging Face Ecosystem},
  author={Li, Ningyuan and Zhao, Yanjie and Wang, Shenao and Wu, Zehao and Wang, Haoyu},
  booktitle={Proceedings of the 16th International Conference on Internetware},
  pages={498--509},
  year={2025}
}

@inproceedings{liu2026art,
  title={The Art of Hide and Seek: Making $\{$Pickle-Based$\}$ Model Supply Chain Poisoning Stealthy Again},
  author={Liu, Tong and Meng, Guozhu and Zhou, Peng and Deng, Zizhuang and Yao, Shuaiyin and Chen, Kai},
  booktitle={35th USENIX Security Symposium (USENIX Security 26)},
  year={2026}
}

@article{liu2025empirical,
  title={An empirical study of vulnerable package dependencies in LLM repositories},
  author={Liu, Shuhan and Hu, Xing and Xia, Xin and Lo, David and Yang, Xiaohu},
  journal={arXiv preprint arXiv:2508.21417},
  year={2025}
}

@article{meiklejohn2025machine,
  title={Machine learning models have a supply chain problem},
  author={Meiklejohn, Sarah and Blauzvern, Hayden and Maruseac, Mihai and Schrock, Spencer and Simon, Laurent and Shumailov, Ilia},
  journal={arXiv preprint arXiv:2505.22778},
  year={2025}
}

@misc{newspicklescan,
  title = {Picklescan Bugs Allow Malicious {{PyTorch}} Models to Evade Scans and Execute Code},
  author = {News, The Hacker},
  journal = {The Hacker News},
  url = {https://thehackernews.com/2025/12/picklescan-bugs-allow-malicious-pytorch.html},
  chapter = {Article}
}

@inproceedings{nguyen2026securing,
  title={Securing the AI supply chain: What can we learn from developer-reported security issues and solutions of AI projects?},
  author={Nguyen, The Anh and Le, Triet Huynh Minh and Babar, M Ali},
  booktitle={Proceedings of the 2026 IEEE/ACM 48th International Conference on Software Engineering},
  pages={1893--1905},
  year={2026}
}

@inproceedings{kholoosi2026software,
  title={Software vulnerability management in the era of artificial intelligence: An industry perspective},
  author={Kholoosi, M Mehdi and Le, Triet Huynh Minh and Babar, M Ali},
  booktitle={Proceedings of the 2026 IEEE/ACM 48th International Conference on Software Engineering},
  pages={1043--1055},
  year={2026}
}

@inproceedings{le2021large,
  title={A large-scale study of security vulnerability support on developer q\&a websites},
  author={Le, Triet Huynh Minh and Croft, Roland and Hin, David and Babar, Muhammad Ali},
  booktitle={Proceedings of the 25th International Conference on Evaluation and Assessment in Software Engineering},
  pages={109--118},
  year={2021}
}

@inproceedings{le2020puminer,
  title={Puminer: Mining security posts from developer question and answer websites with pu learning},
  author={Le, Triet Huynh Minh and Hin, David and Croft, Roland and Babar, M Ali},
  booktitle={Proceedings of the 17th International Conference on Mining Software Repositories},
  pages={350--361},
  year={2020}
}

@inproceedings{nguyen2024automated,
  title={Automated code-centric software vulnerability assessment: How far are we? an empirical study in c/c++},
  author={Nguyen, Anh The and Le, Triet Huynh Minh and Babar, M Ali},
  booktitle={Proceedings of the 18th ACM/IEEE International Symposium on Empirical Software Engineering and Measurement},
  pages={72--83},
  year={2024}
}

@article{le2022survey,
  title={A survey on data-driven software vulnerability assessment and prioritization},
  author={Le, Triet Huynh Minh and Chen, Huaming and Babar, M Ali},
  journal={ACM Computing Surveys},
  volume={55},
  number={5},
  pages={1--39},
  year={2022},
  publisher={ACM New York, NY}
}

@inproceedings{ye2025llmsecconfig,
  title={Llmsecconfig: An llm-based approach for fixing software container misconfigurations},
  author={Ye, Ziyang and Le, Triet Huynh Minh and Babar, M Ali},
  booktitle={2025 IEEE/ACM 22nd International Conference on Mining Software Repositories (MSR)},
  pages={629--641},
  year={2025},
  organization={IEEE}
}

@inproceedings{le2021deepcva,
  title={Deepcva: Automated commit-level vulnerability assessment with deep multi-task learning},
  author={Le, Triet Huynh Minh and Hin, David and Croft, Roland and Babar, M Ali},
  booktitle={2021 36th IEEE/ACM International Conference on Automated Software Engineering (ASE)},
  pages={717--729},
  year={2021},
  organization={IEEE}
}

@inproceedings{le2022use,
  title={On the use of fine-grained vulnerable code statements for software vulnerability assessment models},
  author={Le, Triet Huynh Minh and Babar, M Ali},
  booktitle={Proceedings of the 19th International Conference on Mining Software Repositories},
  pages={621--633},
  year={2022}
}

@article{le2020deep,
  title={Deep learning for source code modeling and generation: Models, applications, and challenges},
  author={Le, Triet Huynh Minh and Chen, Hao and Babar, Muhammad Ali},
  journal={ACM Computing Surveys (CSUR)},
  volume={53},
  number={3},
  pages={1--38},
  year={2020},
  publisher={ACM New York, NY, USA}
}

@article{arani2024systematic,
  title={Systematic literature review on application of learning-based approaches in continuous integration},
  author={Arani, Ali Kazemi and Le, Triet Huynh Minh and Zahedi, Mansooreh and Babar, M Ali},
  journal={IEEE Access},
  volume={12},
  pages={135419--135450},
  year={2024},
  publisher={IEEE}
}

@inproceedings{arani2023sok,
  title={Sok: Machine learning for continuous integration},
  author={Arani, Ali Kazemi and Zahedi, Mansooreh and Le, Triet Huynh Minh and Babar, Muhammad Ali},
  booktitle={2023 IEEE/ACM International Workshop on Cloud Intelligence \& AIOps (AIOps)},
  pages={8--13},
  year={2023},
  organization={IEEE}
}

@inproceedings{le2019automated,
  title={Automated software vulnerability assessment with concept drift},
  author={Le, Triet Huynh Minh and Sabir, Bushra and Babar, Muhammad Ali},
  booktitle={2019 IEEE/ACM 16th International Conference on Mining Software Repositories (MSR)},
  pages={371--382},
  year={2019},
  organization={IEEE}
}

@article{le2022towards,
  title={Towards an improved understanding of software vulnerability assessment using data-driven approaches},
  author={Le, Triet Huynh Minh},
  journal={arXiv preprint arXiv:2207.11708},
  year={2022}
}

@inproceedings{nguyen2025toward,
  title={Toward realistic evaluations of just-in-time vulnerability prediction},
  author={Nguyen, Duong and Le-Cong, Thanh and Le, Triet Huynh Minh and Babar, M Ali and Huynh, Quyet-Thang},
  booktitle={2025 IEEE International Conference on Software Maintenance and Evolution (ICSME)},
  pages={1--10},
  year={2025},
  organization={IEEE}
}

@inproceedings{le2024latent,
  title={Are latent vulnerabilities hidden gems for software vulnerability prediction? an empirical study},
  author={Le, Triet Huynh Minh and Du, Xiaoning and Babar, M Ali},
  booktitle={Proceedings of the 21st International Conference on Mining Software Repositories},
  pages={716--727},
  year={2024}
}

@inproceedings{le2024automatic,
  title={Automatic data labeling for software vulnerability prediction models: How far are we?},
  author={Le, Triet Huynh Minh and Babar, Muhammad Ali},
  booktitle={Proceedings of the 18th ACM/IEEE International Symposium on Empirical Software Engineering and Measurement},
  pages={131--142},
  year={2024}
}

@inproceedings{le2024software,
  title={Software vulnerability prediction in low-resource languages: An empirical study of codebert and chatgpt},
  author={Le, Triet Huynh Minh and Babar, M Ali and Thai, Tung Hoang},
  booktitle={Proceedings of the 28th International Conference on Evaluation and Assessment in Software Engineering},
  pages={679--685},
  year={2024}
}

@article{patel2026towards,
  title={Towards Secure MLOps: Surveying Attacks, Mitigation Strategies, and Research Challenges},
  author={Patel, Raj and Tripathi, Himanshu and Stone, Jasper and Amiri Golilarz, Noorbakhsh and Mittal, Sudip and Rahimi, Shahram and Chaudhary, Vini},
  journal={ACM Transactions on AI Security and Privacy},
  year={2026},
  publisher={ACM New York, NY}
}

@article{siddiq2026empirical,
  title={An empirical study on remote code execution in machine learning model hosting ecosystems},
  author={Siddiq, Mohammed Latif and Romel, Tanzim Hossain and Sekerak, Natalie and Casey, Beatrice and Santos, Joanna},
  journal={arXiv preprint arXiv:2601.14163},
  year={2026}
}

@misc{trailofbitspublications,
  title = {Publications/Reviews/2023-03-Eleutherai-Huggingface-Safetensors-Securityreview.Pdf at Master {$\cdot$} Trailofbits/Publications},
  author = {{Trailofbits}},
  journal = {GitHub},
  url = {https://github.com/trailofbits/publications/blob/master/reviews/2023-03-eleutherai-huggingface-safetensors-securityreview.pdf}
}

@inproceedings{yamaguchi2014modeling,
  title = {Modeling and Discovering Vulnerabilities with Code Property Graphs},
  booktitle = {2014 {{IEEE Symposium}} on {{Security}} and {{Privacy}}},
  author = {Yamaguchi, Fabian and Golde, Nico and Arp, Daniel and Rieck, Konrad},
  year = 2014,
  pages = {590--604},
  publisher = {IEEE},
  doi = {10.1109/SP.2014.44}
}

@inproceedings{yan2002gspan,
  title = {{{gSpan}}: Graph-Based Substructure Pattern Mining},
  booktitle = {2002 {{IEEE International Conference}} on {{Data Mining}}, 2002. {{Proceedings}}.},
  author = {Yan, Xifeng and Han, Jiawei},
  year = 2002,
  pages = {721--724},
  publisher = {IEEE Comput. Soc},
  doi = {10.1109/ICDM.2002.1184038}
}

@article{yasmin2026software,
  title={Software dependencies 2.0: An empirical study of reuse and integration of pre-trained models in open-source projects},
  author={Yasmin, Jerin and Jiang, Wenxin and Davis, James C and Tian, Yuan},
  journal={Empirical Software Engineering},
  volume={31},
  number={6},
  pages={173},
  year={2026},
  publisher={Springer}
}

@inproceedings{zhu2025my,
  title={My model is malware to you: Transforming ai models into malware by abusing tensorflow apis},
  author={Zhu, Ruofan and Chen, Ganhao and Shen, Wenbo and Xie, Xiaofei and Chang, Rui},
  booktitle={2025 IEEE Symposium on Security and Privacy (SP)},
  pages={486--503},
  year={2025},
  organization={IEEE}
}

@inproceedings{zhao2024models,
  title={Models are codes: Towards measuring malicious code poisoning attacks on pre-trained model hubs},
  author={Zhao, Jian and Wang, Shenao and Zhao, Yanjie and Hou, Xinyi and Wang, Kailong and Gao, Peiming and Zhang, Yuanchao and Wei, Chen and Wang, Haoyu},
  booktitle={Proceedings of the 39th IEEE/ACM international conference on automated software engineering},
  pages={2087--2098},
  year={2024}
}

@misc{protectai_insights,
  title        = {AI Supply Chain Security Insights},
  author       = {{Protect AI}},
  year         = {2026},
  howpublished = {\url{https://protectai.com/insights}},
}

@misc{huggingface_protectai_security,
  author       = {{Hugging Face}},
  title        = {Protect AI Security Scanning on the Hugging Face Hub},
  year         = {2026},
  howpublished = {\url{https://huggingface.co/docs/hub/en/security-protectai}},
}

@misc{protectai_knowledge_base,
  author       = {{Protect AI}},
  title        = {AI Supply Chain Security Knowledge Base},
  year         = {2026},
  howpublished = {\url{https://protectai.com/insights/knowledge-base}},
}

@misc{joern_tool,
  author       = {{Joern Project}},
  title        = {Joern: A Platform for Code Property Graphs},
  year         = {2026},
  howpublished = {\url{https://joern.io}},
}

@misc{joern_cpg_slicing,
  author       = {{Joern Project}},
  title        = {CPG Slicing},
  year         = {2026},
  howpublished = {\url{https://docs.joern.io/cpg-slicing/}},
}

@misc{joern_cpg_spec,
  author       = {{Joern Team}},
  title        = {Code Property Graph Specification},
  year         = {2023},
  howpublished = {\url{https://cpg.joern.io/\#schema-ref-cfg}},
}

@misc{networkx,
  author       = {{NetworkX Developers}},
  title        = {NetworkX: Network Analysis in Python},
  year         = {2024},
  howpublished = {\url{https://networkx.org}},
}

@inproceedings{fournier2016spmf,
  title={The SPMF open-source data mining library version 2},
  author={Fournier-Viger, Philippe and Lin, Jerry Chun-Wei and Gomariz, Antonio and Gueniche, Ted and Soltani, Azadeh and Deng, Zhihong and Lam, Hoang Thanh},
  booktitle={Joint European conference on machine learning and knowledge discovery in databases},
  pages={36--40},
  year={2016},
  organization={Springer}
}

@inproceedings{fournier2019tkg,
  title={Tkg: Efficient mining of top-k frequent subgraphs},
  author={Fournier-Viger, Philippe and Cheng, Chao and Lin, Jerry Chun-Wei and Yun, Unil and Kiran, R Uday},
  booktitle={International Conference on Big Data Analytics},
  pages={209--226},
  year={2019},
  organization={Springer}
}

@misc{huggingface_pai_2025,
  title        = {4M Models Scanned: Protect AI + Hugging Face 6 Months In},
  author       = {{Hugging Face}},
  howpublished = {\url{https://huggingface.co/blog/pai-6-month}},
  year         = {2025}
}

@misc{huggingface_docs,
  author       = {{Hugging Face}},
  title        = {Hugging Face Documentation},
  year         = {2026},
  howpublished = {\url{https://huggingface.co/docs}},
}

@online{huggingface_pickle_scanning,
  title        = {Pickle Scanning},
  author       = {{Hugging Face}},
  year         = {2026},
  url          = {https://huggingface.co/docs/hub/security-pickle#pickle-scanning},
  organization = {Hugging Face}
}

@online{python_pickle_module,
  title        = {pickle — Python object serialization},
  author       = {{Python Software Foundation}},
  year         = {2026},
  url          = {https://docs.python.org/3/library/pickle.html},
  organization = {Python.org}
}

@inproceedings{xiao2018security,
  title={Security risks in deep learning implementations},
  author={Xiao, Qixue and Li, Kang and Zhang, Deyue and Xu, Weilin},
  booktitle={2018 IEEE Security and privacy workshops (SPW)},
  pages={123--128},
  year={2018},
  organization={IEEE}
}

@online{jfrog_model_threats,
  title        = {Model Threats},
  author       = {{JFrog Security Research}},
  year         = {2026},
  url          = {https://research.jfrog.com/model-threats/},
  organization = {JFrog Ltd.}
}

@online{activepieces_text_classification_action,
  title        = {Text Classification Action for Hugging Face (Activepieces)},
  author       = {{Activepieces}},
  year         = {2026},
  url          = {https://github.com/activepieces/activepieces/blob/32d00da/packages/pieces/community/hugging-face/src/lib/actions/text-classification.ts},
  organization = {Activepieces}
}

@online{activepieces_repo,
  title        = {Activepieces: Open-source AI automation and workflow platform},
  author       = {{Activepieces}},
  year         = {2026},
  url          = {https://github.com/activepieces/activepieces},
  organization = {Activepieces}
}

@online{yuvarajareddy_youtube_sentiment_space,
  title        = {YouTube Comments Sentiment},
  author       = {{yuvarajareddy001}},
  year         = {2026},
  url          = {https://huggingface.co/spaces/yuvarajareddy001/youtube_comments_sentiment},
  organization = {Hugging Face}
}

@inproceedings{jiang2024app,
  title={App-miner: Detecting api misuses via automatically mining api path patterns},
  author={Jiang, Jiasheng and Wu, Jingzheng and Ling, Xiang and Luo, Tianyue and Qu, Sheng and Wu, Yanjun},
  booktitle={2024 IEEE Symposium on Security and Privacy (SP)},
  pages={4034--4052},
  year={2024},
  organization={IEEE}
}

@inproceedings{cheng2009identifying,
  title={Identifying bug signatures using discriminative graph mining},
  author={Cheng, Hong and Lo, David and Zhou, Yang and Wang, Xiaoyin and Yan, Xifeng},
  booktitle={Proceedings of the eighteenth international symposium on Software testing and analysis},
  pages={141--152},
  year={2009}
}

@inproceedings{wi2022hiddencpg,
  title={HiddenCPG: large-scale vulnerable clone detection using subgraph isomorphism of code property graphs},
  author={Wi, Seongil and Woo, Sijae and Whang, Joyce Jiyoung and Son, Sooel},
  booktitle={Proceedings of the ACM web conference 2022},
  pages={755--766},
  year={2022}
}

@article{milo2002network,
  title={Network motifs: simple building blocks of complex networks},
  author={Milo, Ron and Shen-Orr, Shai and Itzkovitz, Shalev and Kashtan, Nadav and Chklovskii, Dmitri and Alon, Uri},
  journal={Science},
  volume={298},
  number={5594},
  pages={824--827},
  year={2002},
  publisher={American Association for the Advancement of Science}
}

@inproceedings{narayanan2012effects,
  title={The effects of different representations on malware motif identification},
  author={Narayanan, Ajit and Chen, Yi and Pang, Shaoning and Tao, Ban},
  booktitle={2012 Eighth International Conference on Computational Intelligence and Security},
  pages={86--90},
  year={2012},
  organization={IEEE}
}

@article{thomas2010margin,
  title={Margin: Maximal frequent subgraph mining},
  author={Thomas, Lini T and Valluri, Satyanarayana R and Karlapalem, Kamalakar},
  journal={ACM Transactions on Knowledge Discovery from Data (TKDD)},
  volume={4},
  number={3},
  pages={1--42},
  year={2010},
  publisher={ACM New York, NY, USA}
}

@inproceedings{hellal2015maximal,
  title={Maximal frequent sub-graph mining for malware detection},
  author={Hellal, Aya and Romdhane, Lotfi Ben},
  booktitle={2015 15th International Conference on Intelligent Systems Design and Applications (ISDA)},
  pages={31--39},
  year={2015},
  organization={IEEE}
}

@inproceedings{dmitriev2006mining,
  title={Mining Generalized Graph Patterns based on User Examples},
  author={Dmitriev, Pavel and Lagoze, Carl},
  booktitle={Sixth International Conference on Data Mining (ICDM'06)},
  pages={857--862},
  year={2006},
  organization={IEEE}
}

@incollection{cheng2010discriminative,
  title={Discriminative frequent pattern-based graph classification},
  author={Cheng, Hong and Yan, Xifeng and Han, Jiawei},
  booktitle={Link Mining: Models, Algorithms, and Applications},
  pages={237--262},
  year={2010},
  publisher={Springer}
}

@article{yamaguchi2015pattern,
  title={Pattern-based methods for vulnerability discovery.},
  author={Yamaguchi, Fabian},
  journal={IT: Information Technology},
  volume={59},
  number={2},
  pages={101},
  year={2017}
}

@misc{huggingface_datasets_loading_methods,
  title        = {Datasets Loading Methods — verification\_mode},
  author       = {{Hugging Face}},
  year         = {2024},
  howpublished = {\url{https://huggingface.co/docs/datasets/v4.8.4/en/package_reference/loading_methods\#datasets.load_dataset.verification_mode}},
}

@misc{pytorch_torch_load_doc,
  author       = {{PyTorch Contributors}},
  title        = {PyTorch Documentation: torch.load},
  year         = {2025},
  url          = {https://docs.pytorch.org/docs/stable/generated/torch.load.html},
}

@misc{pytorch_weights_only_security,
  author       = {{PyTorch Contributors}},
  title        = {PyTorch Serialization Notes: weights\_only Security},
  year         = {2025},
  url          = {https://docs.pytorch.org/docs/stable/notes/serialization.html#weights-only-security},
}

@misc{joblib_load,
  author       = {{Joblib Developers}},
  title        = {joblib.load --- joblib documentation},
  year         = {2026},
  url          = {https://joblib.readthedocs.io/en/latest/generated/joblib.load.html},
}

@article{casey2024large,
  title={A large-scale exploit instrumentation study of AI/ML supply chain attacks in hugging face models},
  author={Casey, Beatrice and Santos, Joanna and Mirakhorli, Mehdi},
  journal={arXiv preprint arXiv:2410.04490},
  year={2024}
}

@inproceedings{kellas2025pickleball,
  title={PickleBall: Secure Deserialization of Pickle-based Machine Learning Models},
  author={Kellas, Andreas D and Christou, Neophytos and Jiang, Wenxin and Li, Penghui and Simon, Laurent and David, Yaniv and Kemerlis, Vasileios P and Davis, James C and Yang, Junfeng},
  booktitle={Proceedings of the 2025 ACM SIGSAC Conference on Computer and Communications Security},
  pages={3341--3355},
  year={2025}
}

@misc{keras_model_saving_loading,
  author       = {{Keras Team}},
  title        = {Whole model saving and loading --- Keras 3 API documentation},
  year         = {2026},
  url          = {https://keras.io/api/models/model_saving_apis/model_saving_and_loading/},
}

@misc{ultralytic_attack,
  author = {Ultralytics},
  title = {Issue 18027},
  howpublished = {\url{https://github.com/ultralytics/ultralytics/issues/18027}},
}

@article{snyk_ultralytics,
  title        = {Ultralytics AI Pwn: Request Supply Chain Attack},
  author       = {Snyk Security Team},
  journal      = {Snyk Blog},
  year         = {2023},
  url          = {https://snyk.io/blog/ultralytics-ai-pwn-request-supply-chain-attack/},
}

@article{hiddenlayer_ultralytics,
  title        = {Ultralytics Python Package Compromise Deploys Cryptominer},
  author       = {Hidden Layer},
  journal      = {Innovation Hub},
  year         = {2023},
  url          = {https://hiddenlayer.com/innovation-hub/ultralytics-python-package-compromise-deploys-cryptominer/},
}

@misc{ultralytics_model_py,
  author       = {{Ultralytics}},
  title        = {Model class implementation in ultralytics/engine/model.py},
  year         = {2026},
  url          = {https://github.com/ultralytics/ultralytics/blob/main/ultralytics/engine/model.py#L123},
}

@article{wang2024mitigating,
  title={Mitigating Downstream Model Risks via Model Provenance},
  author={Wang, Keyu and Iranzad, Abdullah Norozi and Schaffter, Scott and Risdal, Meg and Precup, Doina and Lebensold, Jonathan},
  journal={arXiv preprint arXiv:2410.02230},
  year={2024}
}

@article{jafari2023dependency,
  title={Dependency practices for vulnerability mitigation},
  author={Jafari, Abbas Javan and Costa, Diego Elias and Abdellatif, Ahmad and Shihab, Emad},
  journal={arXiv preprint arXiv:2310.07847},
  year={2023}
}

@article{kula2018developers,
  title={Do developers update their library dependencies? An empirical study on the impact of security advisories on library migration},
  author={Kula, Raula Gaikovina and German, Daniel M and Ouni, Ali and Ishio, Takashi and Inoue, Katsuro},
  journal={Empirical Software Engineering},
  volume={23},
  number={1},
  pages={384--417},
  year={2018},
  publisher={Springer}
}

@inproceedings{digregorio2026security,
  title={On the (In) Security of Loading Machine Learning Models},
  author={Digregorio, Gabriele and Di Gennaro, Marco and Zanero, Stefano and Longari, Stefano and Carminati, Michele and others},
  booktitle={Proceedings of 2026 IEEE Symposium on Security and Privacy},
  pages={1--18},
  year={2026},
  organization={IEEE}
}

@misc{huggingface_hub_docs,
  author       = {{Hugging Face}},
  title        = {Hugging Face Hub Documentation},
  year         = {2026},
  url          = {https://huggingface.co/docs/hub/index},
}

@inproceedings{li2022mining,
  title={Mining node. js vulnerabilities via object dependence graph and query},
  author={Li, Song and Kang, Mingqing and Hou, Jianwei and Cao, Yinzhi},
  booktitle={31st USENIX Security Symposium (USENIX Security 22)},
  pages={143--160},
  year={2022}
}

@inproceedings{khodayari2021jaw,
  title={$\{$JAW$\}$: Studying client-side $\{$CSRF$\}$ with hybrid property graphs and declarative traversals},
  author={Khodayari, Soheil and Pellegrino, Giancarlo},
  booktitle={30th usenix security symposium (usenix security 21)},
  pages={2525--2542},
  year={2021}
}

@article{song2019inferring,
  title={Inferring patterns for taint-style vulnerabilities with security patches},
  author={Song, Tingyu and Li, Xiaohong and Feng, Zhiyong and Xu, Guangquan},
  journal={IEEE Access},
  volume={7},
  pages={52339--52349},
  year={2019},
  publisher={IEEE}
}

@misc{huggingface_model_cards,
  title        = {Model Cards},
  author       = {{Hugging Face}},
  year         = {2024},
  howpublished = {\url{https://huggingface.co/docs/hub/model-cards}},
}

@inproceedings{mitchell2019model,
  title={Model cards for model reporting},
  author={Mitchell, Margaret and Wu, Simone and Zaldivar, Andrew and Barnes, Parker and Vasserman, Lucy and Hutchinson, Ben and Spitzer, Elena and Raji, Inioluwa Deborah and Gebru, Timnit},
  booktitle={Proceedings of the conference on fairness, accountability, and transparency},
  pages={220--229},
  year={2019}
}

@article{suneja2020learning,
  title={Learning to map source code to software vulnerability using code-as-a-graph},
  author={Suneja, Sahil and Zheng, Yunhui and Zhuang, Yufan and Laredo, Jim and Morari, Alessandro},
  journal={arXiv preprint arXiv:2006.08614},
  year={2020}
}

@inproceedings{bowman2020vgraph,
  title={VGRAPH: A robust vulnerable code clone detection system using code property triplets},
  author={Bowman, Benjamin and Huang, H Howie},
  booktitle={2020 IEEE European Symposium on Security and Privacy (EuroS\&P)},
  pages={53--69},
  year={2020},
  organization={IEEE}
}

@inproceedings{foggia2001improved,
  title={An improved algorithm for matching large graphs},
  author={Cordella, Luigi Pietro and Foggia, Pasquale and Sansone, Carlo and Vento, Mario and others},
  booktitle={3rd IAPR-TC15 workshop on graph-based representations in pattern recognition},
  pages={149--159},
  year={2001}
}

@article{cordella2004sub,
  title={A (sub) graph isomorphism algorithm for matching large graphs},
  author={Cordella, Luigi P and Foggia, Pasquale and Sansone, Carlo and Vento, Mario},
  journal={IEEE transactions on pattern analysis and machine intelligence},
  volume={26},
  number={10},
  pages={1367--1372},
  year={2004},
  publisher={IEEE}
}

@misc{replication_package_vul_model_usage,
    title = "Replication Package",
    author = "",
    year=2026,
    howpublished = {\url{https://figshare.com/s/b24b365e34902cb63ddf}},
}

@misc{spark-nlp-github,
  author       = {{John Snow Labs}},
  title        = {Spark NLP: State of the Art Natural Language Processing Library},
  year         = {2025},
  howpublished = {\url{https://github.com/JohnSnowLabs/spark-nlp}},
}

@misc{lxyuan2023distilbert-multilingual-sentiment,
  author       = {lxyuan},
  title        = {distilbert-base-multilingual-cased-sentiments-student},
  year         = {2023},
  howpublished = {\url{https://huggingface.co/lxyuan/distilbert-base-multilingual-cased-sentiments-student}},
}

@misc{protectai2024distilbert-sentiment-overview,
  author       = {{Protect AI}},
  title        = {distilbert-base-multilingual-cased-sentiments-student: Model Overview},
  year         = {2024},
  howpublished = {\url{https://protectai.com/insights/models/lxyuan/distilbert-base-multilingual-cased-sentiments-student/cf991100d706c13c0a080c097134c05b7f436c45/overview}},
}

@inproceedings{bommasani2024ecosystem,
  title={Ecosystem graphs: Documenting the foundation model supply chain},
  author={Bommasani, Rishi and Soylu, Dilara and Liao, Thomas I and Creel, Kathleen A and Liang, Percy},
  booktitle={Proceedings of the AAAI/ACM Conference on AI, Ethics, and Society},
  volume={7},
  number={1},
  pages={196--209},
  year={2024}
}

@article{stalnaker2025empirical,
  title={An Empirical Analysis of Machine Learning Model and Dataset Documentation, Supply Chain, and Licensing Challenges on Hugging Face},
  author={Stalnaker, Trevor and Wintersgill, Nathan and Chaparro, Oscar and Heymann, Laura A and Di Penta, Massimiliano and German, Daniel M and Poshyvanyk, Denys},
  journal={ACM Transactions on Software Engineering and Methodology},
  year={2025},
  publisher={ACM New York, NY}
}

@inproceedings{decan2018impact,
  title={On the impact of security vulnerabilities in the npm package dependency network},
  author={Decan, Alexandre and Mens, Tom and Constantinou, Eleni},
  booktitle={Proceedings of the 15th international conference on mining software repositories},
  pages={181--191},
  year={2018}
}

@inproceedings{zimmermann2019small,
  title={Small world with high risks: A study of security threats in the npm ecosystem},
  author={Zimmermann, Markus and Staicu, Cristian-Alexandru and Tenny, Cam and Pradel, Michael},
  booktitle={28th USENIX Security symposium (USENIX security 19)},
  pages={995--1010},
  year={2019}
}

@inproceedings{jiang2024peatmoss,
  title={Peatmoss: A dataset and initial analysis of pre-trained models in open-source software},
  author={Jiang, Wenxin and Yasmin, Jerin and Jones, Jason and Synovic, Nicholas and Kuo, Jiashen and Bielanski, Nathaniel and Tian, Yuan and Thiruvathukal, George K and Davis, James C},
  booktitle={Proceedings of the 21st International Conference on Mining Software Repositories},
  pages={431--443},
  year={2024}
}

@misc{huggingface_hf_hub_download,
  author       = {{Hugging Face}},
  title        = {hf\_hub\_download},
  howpublished = {\url{https://huggingface.co/docs/huggingface_hub/v1.10.1/en/package_reference/file_download#huggingface_hub.hf_hub_download}},
  year         = {2023}
}

@inproceedings{stalnaker2025understanding,
  title={Understanding and Supporting the ML Supply Chain Through ML Bill of Materials},
  author={Stalnaker, Trevor},
  booktitle={2025 IEEE/ACM 47th International Conference on Software Engineering: Companion Proceedings (ICSE-Companion)},
  pages={1--3},
  year={2025},
  organization={IEEE}
}

@misc{nist2024securing,
  title = {Securing AI Ecosystems: The Critical Role of AIBOM},
  author = {{National Institute of Standards and Technology (NIST)}},
  year = {2024},
  howpublished = {\url{https://csrc.nist.gov/presentations/2024/securing-ai-ecosystems-the-critical-role-of-aibom}},
}

@article{radanliev2024capability,
  title={Capability hardware enhanced instructions and artificial intelligence bill of materials in trustworthy artificial intelligence systems: Analyzing cybersecurity threats, exploits, and vulnerabilities in new software bills of materials with artificial intelligence},
  author={Radanliev, Petar and Santos, Omar and Brandon-Jones, Alistair},
  journal={The Journal of Defense Modeling and Simulation},
  pages={15485129241267919},
  year={2024},
  publisher={SAGE Publications Sage UK: London, England}
}

@inproceedings{xia2024trust,
  title={Trust in software supply chains: Blockchain-enabled sbom and the aibom future},
  author={Xia, Boming and Zhang, Dawen and Liu, Yue and Lu, Qinghua and Xing, Zhenchang and Zhu, Liming},
  booktitle={Proceedings of the 2024 ACM/IEEE 4th International Workshop on Engineering and Cybersecurity of Critical Systems (EnCyCriS) and 2024 IEEE/ACM Second International Workshop on Software Vulnerability},
  pages={12--19},
  year={2024}
}

@misc{cisa_vulnerability_exploitability,
  author       = {{Cybersecurity and Infrastructure Security Agency (CISA)}},
  title        = {Minimum Requirements for Vulnerability Exploitability eXchange (VEX)},
  year         = {2023},
  howpublished = {\url{https://www.cisa.gov/resources-tools/resources/minimum-requirements-vulnerability-exploitability-exchange-vex}},
}

@misc{wnic00_distilbert_p1,
  author       = {wnic00},
  title        = {distilbert-p1},
  year         = {2026},
  howpublished = {\url{https://huggingface.co/wnic00/distilbert-p1}},
}

\appendices 

\section{Ethical Considerations}
Our research involves the following stakeholders: (1)~\textit{open-source developers} whose repositories were retrieved and analyzed; (2)~\textit{model authors} whose artifacts on HuggingFace are identified as vulnerable and traced; 
(3)~\textit{platform operators}, specifically HuggingFace and GitHub, whose infrastructure and public APIs were queried; (4)~\textit{the security research community and practitioners} who may act on the findings.
All stakeholders may benefit from improved AI supply chain security.

\textbf{Ethical Principles}:
We considered these principles:

\noindent\textit{Respect for Persons and Projects:}
Our study does not involve human subjects.
All data was collected from publicly accessible sources: the GitHub Code Search API, the HuggingFace Hub API, and the ProtectAI Insights database.
Repositories and model cards analyzed in this study were published openly by their authors.
We only mention specific, publicly available instances to illustrate patterns rather than to assign faults to specific developers.

\noindent\textit{Beneficence.}
The primary goal of this research is to reduce risk across the AI supply chain by showing a gap between artifact-level security advisories and application-level integration practices. 
The findings benefit developers by providing concrete evidence of structural integration risks and actionable guidance for mitigation; framework maintainers by identifying specific API defaults and parameter configurations that are systematically bypassed or left implicit; platform operators by demonstrating the limitations of hub-level advisories in isolation and motivating cross-layer provenance tooling; and the security research community by introducing a data-driven CPG and FSM methodology applicable to future AI supply chain analysis.

\noindent\textit{Respect for Law, Justice, and Public Interest.}
We utilized multiple APIs and data sources to ensure our findings are representative of the ecosystem.
We adhered to the terms of service of these sources.

\textrevision{\textbf{Dual-Use Considerations:}
Our analysis analyzes the adoption, propagation, and recurring integration patterns by tracing the use of models that have already been publicly flagged for security vulnerabilities.
These findings have potential dual-use implications: defenders can use them to develop static-analysis rules, auditing tools, and improved security practices, while malicious actors could potentially use the same information to identify security-relevant integration patterns or attack surfaces. 
We assessed this risk in determining the appropriate level of disclosure.
The vulnerabilities analyzed in this study are publicly disclosed and cataloged in the ProtectAI Insights database and public HuggingFace security advisories; we do not introduce zero-day vulnerabilities or previously undisclosed vulnerabilities. 
Furthermore, our structural motifs and representative code snippets are presented as generalized abstractions, using normalized labels and wildcard leaves to illustrate recurring program structures.
We do not provide functional exploit payloads, attack scripts, or instructions for exploiting the identified patterns.}

\textbf{Mitigation for potential harms:} 
\textrevision{Potential harms of this study include possible reputational harm to model publishers, maintainers, or platforms if findings are misread as proof of maliciousness, increased awareness of ecosystem weaknesses, and potential misuse of the findings to identify additional attack surfaces.}
To mitigate this, we use publicly available data that are already publicly flagged by HuggingFace's own security advisory and documented in ProtectAI's knowledge base, thus vulnerabilities are already disclosed and the information we present does not constitute new disclosures.
We describe propagation pathways at a conceptual level sufficient to motivate the security concerns without providing exploit instructions or identifying previously undisclosed attack surfaces.
Furthermore, we interpret security flags and advisories as risk indicators, not as definitive proof of exploitability or malicious intent.

\textbf{Decision to publish:} 
We decided to proceed because the benefits outweighed the harms.
The study addresses a growing and under-studied security problem in model ecosystems and can support better provenance tooling, downstream auditing, and safer integration practices.
We decided to publish to address the security blind spots in the AI supply chain, and can help platforms, framework maintainers, and downstream developers improve defenses.

\newpage

\section{Meta-Review}

The following meta-review was prepared by the program committee for the 2027
IEEE Symposium on Security and Privacy (S\&P) as part of the review process as
detailed in the call for papers.

\subsection{Summary}
This paper presents an empirical study of adoption, propagation, and usage patterns of vulnerable models in the AI supply chain. The authors analyze 16,977 vulnerable HuggingFace models identified by InsightDB and trace their usage across GitHub repositories, software packages, fine-tuning workflows, and deployed applications. To understand how these models are integrated into software systems, the authors combine Code Property Graph (CPG) analysis with Frequent Subgraph Mining (FSM) to identify recurring usage patterns.

\subsection{Scientific Contributions}
\begin{itemize}
\item Independent Confirmation of Important Results with Limited Prior Research
\item Provides a New Data Set For Public Use
\item Creates a New Tool to Enable Future Science
\item Addresses a Long-Known Issue
\item Identifies an Impactful Vulnerability
\item Provides a Valuable Step Forward in an Established Field
\end{itemize}

\subsection{Reasons for Acceptance}
\begin{enumerate}
\item This paper makes a valuable step forward in AI supply-chain security by empirically showing how models already flagged by security scanners are adopted, propagated, and integrated into downstream software systems, offering a useful perspective that complements prior work focused primarily on model registries.
\end{enumerate}

\subsection{Noteworthy Concerns}
\begin{enumerate}
\item This paper uses the FSM-based methodology, which is not sufficiently evaluated against simpler alternatives or with respect to its claimed reduction in manual effort.
\item The authors conducted rule-based vulnerability-usage analysis where the rules used to flag vulnerable usages appear similar to the patterns already used in vulnerable model detection methodology. Furthermore, the paper provides no evidence of the exploitability of the identified vulnerable model usages. The results should be interpreted in the context of these limitations.
\end{enumerate}

\end{document}